\documentclass[aps,prb,twocolumn,showpacs,10pt,floatfix,superscriptaddress,floatfix]{revtex4-1}
\usepackage{braket}
\usepackage{dsfont}
\usepackage{amsfonts}
\usepackage{amsthm}
\usepackage{amssymb}
\usepackage{amsmath,amscd}
\usepackage{mathtools}
\usepackage{rotating}
\usepackage{color}
\usepackage{framed}
\usepackage{pbox}
\usepackage{xfrac}
\usepackage{extarrows}
\usepackage{enumerate}
\usepackage{graphicx}
\usepackage{float}
\usepackage{afterpage}
\usepackage{epigraph}
\usepackage{comment}
\usepackage[hidelinks]{hyperref}
\usepackage{mathrsfs}
\usepackage{tikz}
\usepackage{circuitikz}
\usepackage{tabularx}
\usepackage{ragged2e}
\usepackage[percent]{overpic}
\usepackage{color,soul}
\soulregister\cite7
\soulregister\onlinecite7
\soulregister\ref7
\soulregister\eqref7

\newcommand{\ve}{\vec e}

\newcommand{\vj}{\vec j}
\newcommand{\vjs}{{\vec j}_{\rm s}}

\newcommand{\vm}{\vec m}

\newcommand{\vmu}{\mbox{\boldmath $\mu$}}
\newcommand{\vmus}{\vmu_{\rm s}}

\newcommand{\vB}{\vec B}
\newcommand{\vC}{{\cal C}}

\newcommand{\vE}{\vec E}

\newcommand{\vZ}{{\cal Z}}
\newcommand{\mI}{{\cal I}}

\newcommand{\Dex}{D_{\rm ex}}

\renewcommand{\vec}[1]{\mathbf{#1}}

\begin{document}
\title{Theory of spin-Hall magnetoresistance in the AC (terahertz) regime}
\author{David A. Reiss}
\affiliation{Dahlem Center for Complex Quantum Systems and Physics Department, Freie Universit\"at Berlin, Arnimallee 14, 14195 Berlin, Germany}
\author{Tobias Kampfrath}
\affiliation{Physics Department, Freie Universit\"at Berlin, Arnimallee 14, 14195 Berlin, Germany, and Department of Physical Chemistry, Fritz-Haber Institut, Faradayweg 4--6, 14195 Berlin}
\author{Piet W. Brouwer}
\affiliation{Dahlem Center for Complex Quantum Systems and Physics Department, Freie Universit\"at Berlin, Arnimallee 14, 14195 Berlin, Germany}

\begin{abstract}
In bilayers consisting of a normal metal (N) with spin-orbit coupling and a ferromagnet (F), the combination of the spin-Hall effect, the spin-transfer torque, and the inverse spin-Hall effect gives a small correction to the in-plane conductivity of N, which is referred to as spin-Hall magnetoresistance (SMR). We here present a theory of the SMR and the associated off-diagonal conductivity corrections for frequencies up to the terahertz regime. We show that the SMR signal has pronounced singularities at the spin-wave frequencies of F, which identifies it as a potential tool for all-electric spectroscopy of magnon modes. A systematic change of the magnitude of the SMR at lower frequencies is associated with the onset of a longitudinal magnonic contribution to spin transport across the F-N interface.
\end{abstract}

\maketitle

\section{Introduction}

In recent years, it was experimentally shown that basic spintronic effects not only operate in the DC and GHz regime, but also in the ultrafast (THz) regime.\cite{Walowski_2016} Examples include the spin-Hall effect (SHE) and its inverse (ISHE)\cite{Kampfrath_2013,Seifert_2018_2} in metals with strong spin-orbit coupling, spin pumping and spin-transfer torque (STT)\cite{Schellekens_2014,Razdolski_2017} at interfaces of non-magnetic metals and ferromagnets, and magnon generation.\cite{Schmidt_2010} Furthermore, ultrafast versions of the spin-Seebeck effect\cite{Kimling_2017,Seifert_2018} and the giant magnetoresistance in ferromagnet$|$normal-metal multilayers\cite{Jin_2015} were demonstrated. Magnetoresistance effects are important for eventual applications, because their utilization on ultrafast time scales has the potential of increasing speeds for the electrical readout of magnetic memories. 

\begin{figure}
\centering
\includegraphics[width=0.45\textwidth]{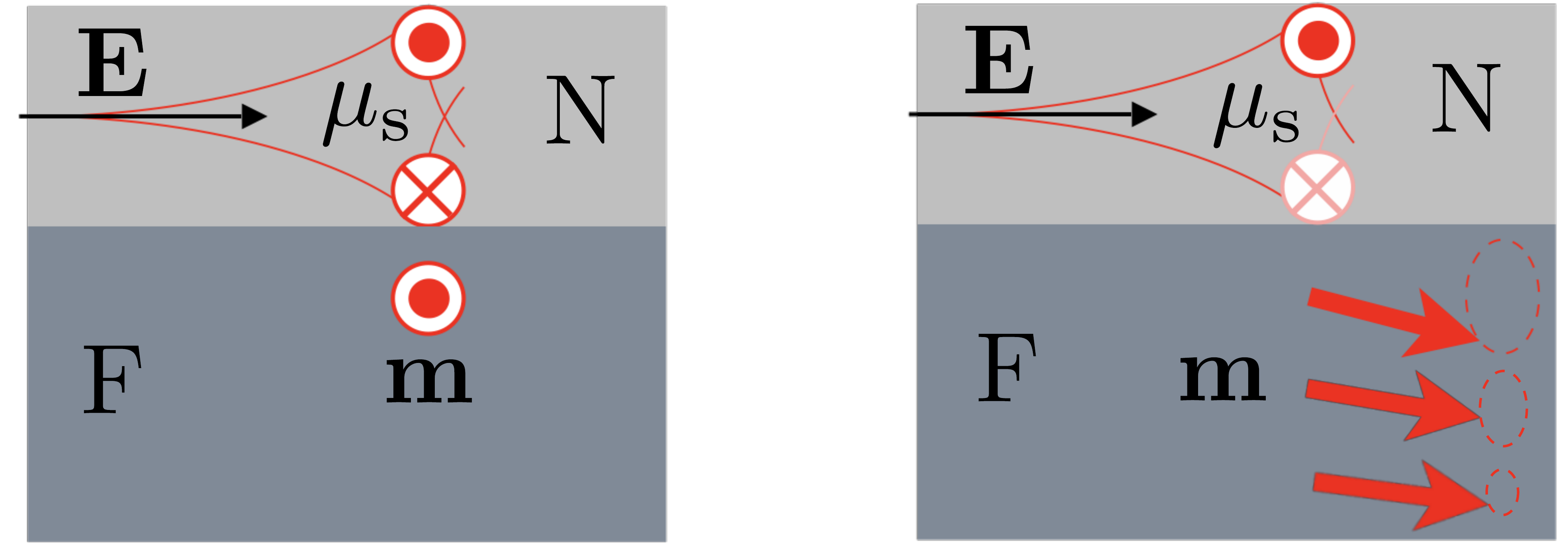}
\caption{Explanation of the SMR in a ferromagnet$|$normal-metal (F$|$N) bilayer: Via the spin-Hall effect, an electric field applied to N creates a spin accumulation at its boundaries. The spin accumulation at the F-N boundary drives a spin current from N into F. This spin current is converted into a charge current in N via the inverse spin-Hall effect. For a time-dependent driving field, additionally magnons (spin waves) can be excited in F. The possibility to coherently excite magnons is what distinguishes the THz SMR from its DC counterpart.\label{fig:schematic}}
\end{figure}

In this article, we consider the ultrafast version of the ``spin-Hall magnetoresistance'' (SMR).\cite{Weiler_2012,Huang_2012,Nakayama_2013,Hahn_2013,Vlietstra_2013,Althammer_2013} The SMR is observed in current-in-plane experiments on bilayers of an insulating or metallic ferromagnet F and a non-magnetic metal N with strong spin-orbit coupling, as shown in Fig. \ref{fig:schematic}. It is a small correction $\delta \sigma^{xx}$ to the diagonal in-plane conductivity ({\em i.e.}, for currents parallel to an applied electric field) that results from a combination of SHE, ISHE, and STT, and that is sensitive to the magnetization direction in the ferromagnetic layer. Measurements of the SMR in the DC regime were found to be in good agreement with theoretical predictions.\cite{Chen_2013,Chen_2016} Lotze {\em et al.} measured the finite-frequency SMR in YIG$|$Pt bilayers up to $3\,{\rm GHz}$ and observed no frequency dependence of $\delta \sigma^{xx}$ within their measurement accuracy.\cite{Lotze_2014} With the present advances in the availability of THz sources,\cite{Fueloep_2019} the experimental investigation of the SMR in the ultrafast regime becomes a realistic possibility.

At zero frequency, the SMR involves a combination of key spintronic phenomena,\cite{Chen_2013,Chen_2016,Zhang_2019} shown schematically in Fig.\ \ref{fig:schematic}: (i) When an electric field $E$ is applied to N, the SHE generates a spin accumulation at the F-N interface. (ii) Via the spin-transfer torque or, in the case of a metallic ferromagnet, via the flow of a spin-polarized electron current, spin angular momentum is transferred between the spin accumulation at the interface and the F layer. (iii) The ISHE converts the spin current through the F-N interface into a charge current in N that flows parallel to the interface. The component of this induced current parallel to the applied electric field corresponds to a change of the diagonal in-plane conductivity $\delta \sigma^{xx}$, whereas the perpendicular current component gives a contribution $\delta \sigma^{xy}$ to the off-diagonal conductivity. The conductivity corrections depend on the magnetization direction, because the amount of angular momentum transferred to F depends on it. The SMR can be distinguished from the anisotropic magnetoresistance in a proximity-induced magnetic N layer by measuring the magnetoresistance for out-of-plane directions of the magnetization.\cite{Chen_2016}

In order to adequately describe the frequency dependence of the SMR, we must decompose the spin transport across the F-N interface into ``longitudinal'' and ``transverse'' contributions, polarized collinear with and perpendicular to the magnetization direction $\vm$, respectively.\cite{Zhang_2019} The difference of the transverse and longitudinal contributions determines the size of the SMR.\cite{Chen_2013,Chen_2016} As we show in detail in this article, both contributions to the SMR depend on frequency, but in different ways. On the one hand, the coherent excitation of spin waves (magnons) causes pronounced singular features in the frequency dependence of the transverse contribution to the SMR, which, with sufficiently high frequency resolution, identifies the SMR as an all-electric spectroscopic probe of magnon modes in F. Collinear spin transport, on the other hand, occurs via the incoherent excitation or annihilation of thermal magnons via spin-flip scattering at the F-N interface\cite{Flipse_2014,Cornelissen_2015,Bender_2015,Cornelissen_2016,Zhang_2019} and, in metallic ferromagnets, via spin-dependent transport of conduction electrons.\cite{Kim_2016} For thin insulating F layers with a long magnon lifetime, longitudinal spin transport is strongly suppressed at zero frequency, as the build-up of an excess density of magnons in F causes a backflow of spin angular momentum from F into N. This compensation mechanism ceases to be effective at finite frequencies, however, which leads to an appreciable decrease of the SMR for typical YIG$|$Pt devices for frequencies in the GHz range and above.

Previous work on the finite-frequency SMR by Chiba, Bauer, and Takahashi\cite{Chiba_2014} theoretically considered transverse spin transport across the F-N interface in a current-in-plane version of the current-induced spin-torque ferromagnetic resonance (FMR)\cite{Liu_2011_2,Kondou_2012,Ganguly_2014} and the non-linear spin-torque diode effect.\cite{Tulapurkar_2005,Sankey_2006} Chiba {\em et al.} considered frequencies close to the FMR frequency $\omega_0$, for which only the uniform mode of the magnetization in the F layer is excited (see also Ref.\ \onlinecite{Guimaraes_2017}). The mechanisms governing transverse spin transport across the F-N interface --- spin torque and spin pumping --- are essentially the same in the THz regime as they are in the lower-GHz regime considered in Ref.\ \onlinecite{Chiba_2014}, which is why this part of our theoretical analysis closely follows Ref.\ \onlinecite{Chiba_2014}. The main difference to Ref.\ \onlinecite{Chiba_2014} is that for driving frequencies in the THz regime the magnetization mode excited by the SHE-induced spin torque is no longer the uniform ferromagnetic-resonance mode, but an acoustic magnon mode, whose wavelength is much shorter than the typical thickness $d_{\rm F}$ of the F layer. Such current-induced coherent magnon excitation was considered theoretically by Sluka\cite{Sluka_2017} and Johansen, Skarsv\aa g, and Brataas\cite{Johansen_2018} for antiferromagnetic layers, in which magnon frequencies are typically higher. We restrict our theory to the linear-response regime and do not consider nonlinear rectification effects responsible for a DC response of driven F$|$N bilayers.\cite{Chiba_2014,Schreier_2015,Sklenar_2015} This is appropriate for the THz regime, because the field amplitudes used in standard THz time-domain spectroscopy are usually too small to induce nonlinear effects.\cite{Jin_2015} 

A theory for the zero-frequency limit that includes both the transverse and the longitudinal contribution to the SMR was considered by Zhang, Bergeret, and Golovach,\cite{Zhang_2019} but without considering the backflow of spin current resulting from a non-equilibrium population of thermal magnons in F. As we show in this article, it is this backflow term that causes a systematic frequency dependence of the SMR in bilayers involving a ferromagnetic insulator with long magnon lifetime. Although, as we show below, the difference between zero-frequency and high-frequency limits does not depend on interface properties and device parameters, the characteristic frequency separating low-frequency and high-frequency regimes depends on these details. The authors of Ref.\ \onlinecite{Lotze_2014} measured the SMR in YIG$|$Pt bilayers in the GHz regime and did not observe an appreciable frequency dependence of the SMR up to approximately $3\,{\rm GHz}$. We attribute the apparent absence of a frequency dependence in this experiment to the presence of the large applied magnetic field, which effectively pinned the magnetization direction, thus shifting the characteristic frequency to a value outside the range accessible in the experiment of Ref. \onlinecite{Lotze_2014}.

Whereas the magnetic field generated by the applied AC current played a significant role if the applied frequencies are close to the ferromagnetic-resonance frequency,\cite{Chiba_2014} the Oersted field only has a minimal effect on the SMR in the THz regime. The reason is that the Oersted field is approximately homogeneous in F such that it can not effectively excite magnon modes at the frequency of the driving field. The same applies to the magnetic field of the electromagnetic wave that drives the SMR at high frequencies. The spin-transfer torque, on the other hand, acts locally at the F-N interface, so that it couples to magnon modes of all wavelengths. For this reason, we will not consider the Oersted field in the main text and, instead, discuss its effect in the appendix.

This article is organized as follows: In Sec.\ \ref{sec:system_notation} we describe the system relevant for the SMR in an F$|$N bilayer geometry and introduce the necessary notation. In Sec.\ \ref{sec:trans+long_SMR}, charge and spin current densities driven by an applied time-dependent electric field are calculated for an F$|$N bilayer with a thickness $d_{\rm N}$ of the N layer much larger than the spin-relaxation length $\lambda_{\rm N}$. This is the geometry relevant for the existing experiments in the low-frequency regime. Following the idea of a ``magneto-electric circuit theory'', the result is formulated in terms of ``spin impedances'' characterizing the N layer, the F layer, and the F-N interface. Separate sets of impedances describe the transverse and longitudinal contributions to the SMR and the associated off-diagonal conductivity corrections. The impedances are calculated from elementary electronic and magnetic transport equations in Sec.\ \ref{sec:impedances}. Specific predictions for the SMR in bilayers of YIG$|$Pt and Fe$|$Au (as prototypes for insulating and metallic ferromagnets) are discussed Sec.\ \ref{sec:experiments}. We conclude in Sec.\ \ref{sec:conclusions}. A discussion of the effect of the Oersted field, of F$|$N bilayers with finite thickness $d_{\rm N} \lesssim \lambda_{\rm N}$ and of F$|$N$|$F trilayers, as well as a theory of the longitudinal magnonic spin transport through the F-N interface with ballistic magnons in F can be found in the appendices.

\section{System and notation}
\label{sec:system_notation}

\begin{figure}
\centering
\includegraphics[width=0.45\textwidth]{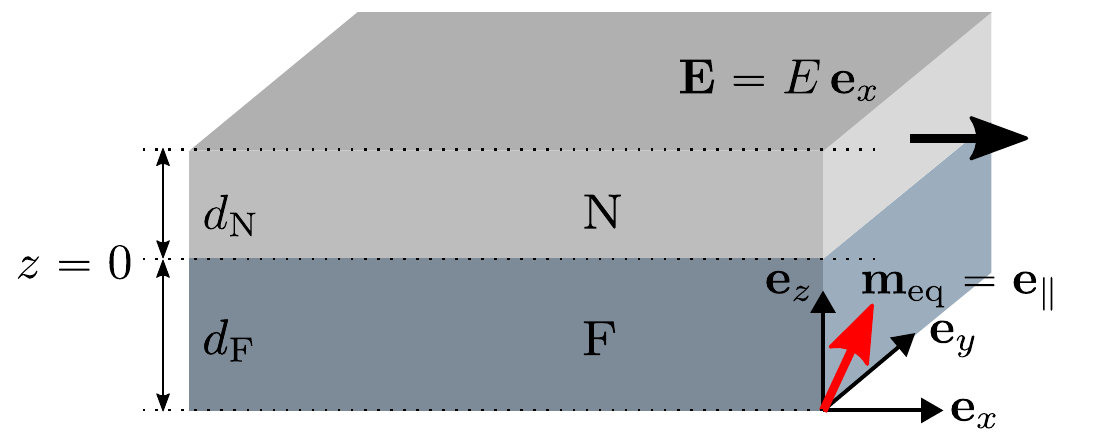}
\caption{F$|$N bilayer with a normal metal of thickness $d_{\rm N}$ and a ferromagnet of thickness $d_{\rm F}$. Coordinates are chosen such that the $xy$ plane is the interface between the F and N layers. The electric field $\vE$ is applied in the $x$ direction.
\label{fig:geometry_notation}}
\end{figure}

We consider the SMR in an F$|$N bilayer geometry, shown schematically in Fig.\ \ref{fig:geometry_notation}. (A discussion of the SMR in an F$|$N$|$F trilayer geometry can be found in App.\ \ref{sec:spin_valves}.)
Following Ref.\ \onlinecite{Chen_2013}, we choose coordinates such that the $z$ direction is perpendicular to the thin films, the normal metal N of thickness $d_{\rm N}$ is located at $0 < z < d_{\rm N}$, and the magnet F at $-d_{\rm F} < z < 0$. A spatially uniform time-dependent electric field $\vE(t) = E(t) \ve_x$ is applied in the $x$ direction.

We assume that the thickness $d_{\rm N}$ of the N layer is much larger than the spin-relaxation length $\lambda_{\rm N}$. In this limit, the the small corrections $\delta \sigma^{xx}$ and $\delta \sigma^{xy}$ to the conductivity of the N layer from the combination of SHE and ISHE are the sum of contributions from the F-N interface at $z=0$ and the N-vacuum interface at $z=d_{\rm N}$. Since the latter does not depend on the magnetization direction $\vm$, for a theory of the SMR it is sufficient to consider the contribution from the F-N interface at $z=0$ only. The case $d_{\rm N} \simeq \lambda_{\rm N}$ is discussed in App.\ \ref{sec:spin_valves}.

The relevant variables for the charge and spin currents in the normal metal N are the charge current densities $j_{\rm c}^{x,y}(z,t)$ in the $x$ and $y$ directions, the spin current density $\vjs^z(z,t)$ flowing in the $z$ direction, and the spin accumulation $\vmus(z,t)$ ($0 < z < d_{\rm N}$) as defined in the following. Throughout we use superscripts to denote spatial directions and subscripts for components associated with the spin direction. We define the spin current density as $j_{{\rm s}z}^z = (\hbar/2e)[j_{{\rm c}\uparrow}^z - j_{{\rm c}\downarrow}^z]$, where $j_{{\rm c}\uparrow,\downarrow}^z$ is the charge current density carried by electrons with spin $\uparrow, \downarrow$ projected onto the $z$ axis, respectively. Similarly, the spin accumulation $\mu_{{\rm s}z} = \mu_{\uparrow} - \mu_{\downarrow}$, where $\mu_{\uparrow,\downarrow}$ is the (electro-)chemical potential for electrons with spin $\uparrow$, $\downarrow$ projected onto the $z$ axis. We use analogous definitions for the components $j_{{\rm s}x}^{z}$, $j_{{\rm s}y}^z$, $\mu_{{\rm s}x}$, and $\mu_{{\rm s}y}$.

The equilibrium direction $\ve_{\parallel}$ of the magnetization of the magnet F is 
\begin{equation}
\label{eq:epolar}
\vm_{\rm eq} \equiv \ve_{\parallel} = m_x \ve_x + m_y \ve_y + m_z \ve_z,
\end{equation}
see Fig. \ref{fig:geometry_notation}. To parameterize the direction perpendicular to $\ve_{\parallel}$, we combine two orthogonal vectors spanning the plane perpendicular to $\ve_{\parallel}$ as the real and imaginary part of a complex unit vector $\ve_{\perp}$ with the property
\begin{equation}
  \ve_{\perp} \times \ve_{\parallel} = + i\, \ve_{\perp},
\end{equation}
which defines $\ve_{\perp}$ up to a phase factor. Anticipating that the $y$ direction plays a special role, as discussed in the next section, a convenient choice is
\begin{align}
  \ve_{\perp} =&\,
\frac{1}{\sqrt{2(m_x^2+m_z^2)}}
\left[(m_x^2+m_z^2) \ve_y 
\right. \nonumber \\ &\, \left. \mbox{}
- (m_z m_y - i m_x) \ve_z 
- (m_x m_y + i m_z) \ve_x \right].
  \label{eq:eperppolar}
\end{align}
(The phase factor of the vector (\ref{eq:eperppolar}) is not defined if $\ve_{\parallel} = \ve_y$, which does not affect the final results.)
To account for the different response to spin excitations collinear with and perpendicular to $\ve_{\parallel}$, we separate the spin accumulation $\vmus$ and the spin current $\vjs^z$ into ``longitudinal'' and ``transverse'' components with respect to the equilibrium magnetization direction $\ve_{\parallel}$,
\begin{align}
  \vmus(z,t) =&\, \mu_{{\rm s}\parallel}(z,t)\ve_{\parallel} 
  + \mu_{{\rm s}\perp}(z,t) \ve_{\perp} + \mu^*_{{\rm s}\perp}(z,t) \ve^*_{\perp}, \nonumber \\
  \vjs^z(z,t) =&\, j_{{\rm s}\parallel}^z(z,t) \ve_{\parallel}
  + j_{{\rm s}\perp}^z(z,t) \ve_{\perp} + j_{{\rm s}\perp}^{z*}(z,t) \ve_{\perp}^*,
  \label{eq:decomposition_long_trans}
\end{align}
where $\mu_{{\rm s}\perp}(z,t)$ and $j_{{\rm s}\perp}^z(z,t)$ are complex variables.

The relevant dynamical variables for the magnet F are the magnetization direction $\vm(z,t)$, the spin current density $\vjs^z(z,t)$ flowing in the $z$ direction, and 
the spin accumulation $\vmus(z,t)$, where $-d_{\rm F} < z < 0$. We consider small deviations of $\vm$ from the equilibrium direction $\ve_{\parallel}$ only, which we parameterize by the complex amplitude $m_{\perp}(z,t)$,
\begin{equation}
  \vm(z,t) = \ve_{\parallel} 
  + m_{\perp}(z,t) \ve_{\perp} + m_{\perp}^*(z,t) \ve_{\perp}^*.
  \label{eq:m1}
\end{equation}
As the exchange field is large, a metallic ferromagnet F can sustain a longitudinal component of the spin accumulation only,
\begin{equation}
  \vmus(z,t) = \mu_{{\rm s}\parallel}(z,t) \ve_{\parallel},\ \ 
  -d_{\rm F} < z < 0,
\end{equation}

\noindent
where we neglect the effect of the SHE in F, because the spin-Hall conductance of common metallic ferromagnets such as Fe, Co and Ni is smaller than that for Pt and Au\cite{Hoffmann_2013,Amin_2019} and does not lead to a frequency dependence.

Performing a Fourier transform to time we write
\begin{equation}
  E(t) = \frac{1}{2 \pi} \int_{-\infty}^{+\infty} d\omega E(\omega) e^{-i \omega t},
\end{equation}
where $E(-\omega) = E^*(\omega)$. The same Fourier representation will be used for all time-dependent quantities introduced above. Note that the transverse amplitudes $j_{{\rm s}\perp}$, $\mu_{{\rm s}\perp}$, and $m_{\perp}$ at frequencies $\omega$ and $-\omega$ need not be complex conjugates of each other because these amplitudes are complex in the time domain. 

\section{SMR for single F-N interface}
\label{sec:trans+long_SMR}

In this section, we state the relations between charge currents, spin currents and spin accumulations in both N and F layers and across the F-N interface. In these relations, ``spin impedances'' appear in a natural way and the results can be formulated as equivalent magnetoelectronic circuit diagrams, similar to electric circuit analysis. In Sec. \ref{sec:impedances}, explicit expressions for the impedances are derived.

To linear order in the applied field and the induced potential gradients, the charge current densities $j_{\rm c}^{x,y}$ and the spin current density $\vjs^z$ in the normal metal N satisfy the characteristic response equations of the SHE and ISHE,\cite{Dyakonov_1971,Dyakonov_1971b,Hirsch_1999,Takahashi_2006}
\begin{align}
  j_{\rm c}^x (z, \omega) =&\, \sigma_{\rm N} E (\omega) - \theta_{\text{SH}} \frac{\sigma_{\rm N}}{2 e} \frac{\partial}{\partial z} \mu_{{\rm s}y} (z, \omega) , \label{eq:jcx} \\
  j_{\rm c}^y (z, \omega) =&\, \theta_{\text{SH}} \frac{\sigma_{\rm N}}{2 e} \frac{\partial}{\partial z} \mu_{{\rm s}x} (z, \omega), \label{eq:jcy} \\
  \vjs^z (z, \omega) =&\, - \frac{\hbar \sigma_{\rm N}}{4 e^2} \frac{\partial}{\partial z} \vmus (z, \omega) - \theta_{\text{SH}} \frac{\hbar \sigma_{\rm N}}{2 e} E (\omega) \ve_{y}.
  \label{eq:jsz}
\end{align}
Here $0 < z < d_{\rm N}$, $\theta_{\rm SH}$ is the spin-Hall angle and $\sigma_{\rm N}$ the conductivity of the N layer. 
Since the thickness $d_{\rm N}$ of the normal metal is assumed to be much larger than its spin-relaxation length $\lambda_{\rm N}$, the spin accumulation near the F-N interface at $z = 0$ does not lead to a spin current for $z$ sufficiently far away from the interface.  Averaging Eqs.\ (\ref{eq:jcx}) and (\ref{eq:jcy}) over $z$, we may then express the ISHE corrections $\delta \bar j_{\rm c}^{x,y}$ to the (effective) current densities associated with the F-N interface: in terms of the spin accumulation $\vmus$ at $z=0$ they read as
\begin{align}
  \label{eq:jcx1}
  \delta \bar j_{\rm c}^{x}(\omega) =&\, \theta_{\rm SH}
     \frac{  \sigma_{\rm N}}{2 e d_{\rm N}} 
     \mu_{{\rm s}y}(z \downarrow 0,\omega), \\
  \label{eq:jcy1}
  \delta \bar j_{\rm c}^{y}(\omega) =&\, - \theta_{\rm SH}
     \frac{\sigma_{\rm N}}{2 e d_{\rm N}} 
    \mu_{{\rm s}x}(z \downarrow 0,\omega) .
\end{align}

To solve for the spin accumulation $\vmus(z \downarrow 0,\omega)$, we observe that, within linear response, Eq.\ \eqref{eq:jsz} implies that $\vmus(z \downarrow 0,\omega)$ must be proportional to \mbox{$\vjs^z(0,\omega)+\theta_{\rm SH} \hbar \sigma_{\rm N} E(\omega) \ve_y/2 e$}. The proportionality relation may be written as
\begin{align}
  \vmus(z \downarrow 0,\omega) &= 
  Z_{\rm N}(\omega) \left[\vjs^z(0,\omega)
    + \theta_{\rm SH} \frac{\hbar \sigma_{\rm N}}{2 e} E(\omega) \ve_y \right],
  \label{eq:ZN1}
\end{align}
where the ``spin impedance'' $Z_{\rm N}(\omega)$ has the dimension of ``area''. We have written $\vjs^z(0,\omega)$ instead of $\vjs^z(z \downarrow 0,\omega)$ because the spin current density is conserved across the F-N interface.

In Eq.\ \eqref{eq:ZN1} we neglect a contribution from the spin-Hanle effect to the magnetoresistance,\cite{Dyakonov2007, Velez_2016} because the direct effect of the applied magnetic field on the spin accumulation in N is typically weak compared to that of the coupling to the ferromagnet F. Inclusion of the spin-Hanle effect into our theory would require the introduction of a spin impedance $Z_{\rm N}(\omega)$ that differs for directions collinear with and perpendicular to the applied magnetic field, but without an additional frequency dependence.

Equation \eqref{eq:ZN1} completely determines the spin accumulation $\vmu_{\rm s}(z \downarrow 0,\omega)$ at the interface to a non-magnetic insulator, since $\vj_{\rm s}^{z}(0,\omega) = 0$ in that case. To find $\vmu_{\rm s}(z \downarrow 0,\omega)$ for an interface with an insulating or metallic magnet F, as is appropriate for the geometry of Fig. \ref{fig:geometry_notation}, we now consider the spin currents through an F-N interface and in F. In the ferromagnet, the spin current $\vj_{\rm s}^z = \vj_{\rm se}^z + \vj_{\rm sm}^z$ has contributions from magnons and conduction electrons. The spin current $\vj_{\rm se}^z = j_{{\rm se}\parallel}^z \ve_{\parallel}$ carried by conduction electrons has a longitudinal component only. Again, within linear response there is a simple proportionality to the (electron) spin accumulation $\mu_{{\rm s}\parallel}(z \uparrow 0,\omega)$ at the magnetic side of the interface, which can be written in a form similar to Eq.\ (\ref{eq:ZN1}),
\begin{align}
  \mu_{{\rm s}\parallel}(z \uparrow 0,\omega) =&\,
  - Z^{\rm e}_{{\rm F}\parallel}(\omega) 
  j_{{\rm se}\parallel}^z(0,\omega).
  \label{eq:ZF1}
\end{align}
An additional equation for $\mu_{{\rm s}\parallel}(z \uparrow 0)$ is found by considering the boundary conditions at the F-N interface, which also take the simple form of a proportionality,\cite{Brataas_2000,Brataas_2001,Tserkovnyak_2005}
\begin{align}
  \mu_{{\rm s}\parallel}(z \downarrow 0,\omega) -
  \mu_{{\rm s}\parallel}(z \uparrow 0,\omega) =&\,
  - Z_{{\rm FN}\parallel}^{\rm e}(\omega) 
  j_{{\rm se}\parallel}^z(0,\omega).
  \label{eq:ZFN0}
\end{align}
As in Eq.\ (\ref{eq:ZN1}), the proportionality constants $Z^{\rm e}_{{\rm F}\parallel}$ and $Z_{{\rm FN}\parallel}^{\rm e}$ have the dimension of ``area''. 

The magnonic spin current $\vj_{\rm sm}^z$ has a coherent transverse component related to the magnetization dynamics as well as an incoherent longitudinal component carried by thermal magnons. In linear response, the equations for the longitudinal magnonic spin current are analogous to Eqs.\ (\ref{eq:ZF1}) and (\ref{eq:ZFN0}) for the electronic spin current,\cite{Flipse_2014,Bender_2015}
\begin{align}
\mu_{{\rm m}}(0,\omega) =&\, - Z_{{\rm F} \parallel}^{\rm m} (\omega) j_{{\rm sm}\parallel}^z(0,\omega),
\label{eq:ZFm2} \\
\mu_{{\rm s}\parallel}(z \downarrow 0,\omega) - \mu_{{\rm m}}(0,\omega) =&\, - Z_{{\rm FN} \parallel}^{\rm m} (\omega) j_{{\rm sm}\parallel}^z(0,\omega), 
\label{eq:ZFN2} 
\end{align}
where $\mu_{{\rm m}}(0,\omega)$ is the chemical potential describing the distribution of thermal magnons in F.\cite{Cornelissen_2016} The transverse component $j_{{\rm sm}\perp}^z$ is proportional to the time derivative of the magnetization,
\begin{align}
  - \hbar \omega m_{\perp}(0,\omega) =&\,
  - Z_{{\rm F}\perp}(\omega) 
    j_{{\rm sm}\perp}^z(0,\omega),
\end{align}
and satisfies the boundary condition\cite{Tserkovnyak_2002_a,Tserkovnyak_2002_b}
\begin{align}
  \mu_{{\rm s}\perp}(z \downarrow 0,\omega) +
  \hbar \omega m_{\perp}(0,\omega) =&\,
  - Z_{{\rm FN}\perp}(\omega) j_{{\rm sm}\perp}^z(0,\omega)
  \label{eq:ZFN1}
\end{align}
at the interface. Again, the proportionality constants $Z_{{\rm FN} \parallel}^{\rm m}$, $Z_{{\rm F} \parallel}^{\rm m}$, $Z_{{\rm F}\perp}$, and $Z_{{\rm FN} \perp}$ have the dimension of ``area''. 
The interface impedances $Z_{{\rm FN}\parallel}^{\rm e}$, $Z_{{\rm FN}\parallel}^{\rm m}$, and $Z_{{\rm FN}\perp}$ may be expressed in terms of the spin-dependent interface conductances $g_{\uparrow\uparrow}$, $g_{\downarrow\downarrow}$ and the spin-mixing conductance $g_{\uparrow\downarrow}$ that are used in the theory of Ref. \onlinecite{Chen_2013,Chen_2016}, see Secs. \ref{subsec:interface_long_el}, \ref{subsec:interface}, and \ref{subsec:ferromagnet_long_mag}.

In the next section we show that of the seven ``spin impedances'' defined in Eqs.\ (\ref{eq:ZN1})--(\ref{eq:ZFN1}) only $Z_{{\rm F}\perp}(\omega)$ and $Z_{{\rm F} \parallel}^{\rm m} (\omega)$ --- associated with the coherent magnon excitation and the non-equilibrium accumulation of magnons, respectively --- have a non-negligible frequency dependence in the THz regime and below. Anticipating this result, we retain the frequency argument for $Z_{{\rm F}\perp}(\omega)$ and $Z_{{\rm F} \parallel}^{\rm m} (\omega)$, but drop it for the five other spin impedances.

Solving the coupled equations (\ref{eq:ZN1})--(\ref{eq:ZFN1}) for the longitudinal and transverse components of the spin current density $\vjs^z$ is straightforward and one finds
\begin{align}
\label{eq:js}
  j^z_{{\rm s}\parallel}(0,\omega) =&\,
  -  \frac{Z_{\rm N}}{Z_{\parallel} (\omega)} \theta_{\rm SH}  \frac{\hbar \sigma_{\rm N}}{2 e}
    E(\omega) \ve_{\parallel} \cdot \ve_y, \nonumber \\
  j^z_{{\rm s}\perp}(0,\omega) =&\,
  - 
  \frac{Z_{\rm N}}{Z_{\perp}(\omega)} \theta_{\rm SH}  \frac{\hbar \sigma_{\rm N}}{2 e}
    E(\omega) \ve_{\perp}^* \cdot \ve_y, 
\end{align}
where we defined
\begin{align}
  Z_{\parallel}(\omega)  =&\ Z_{\rm N} + \left[ \frac{1}{Z_{{\rm FN}\parallel}^{\rm m} + Z_{{\rm F} \parallel}^{\rm m} (\omega)} + \frac{1}{Z_{{\rm FN}\parallel}^{\rm e} + Z^{\rm e}_{{\rm F}\parallel}} \right]^{-1}, \nonumber \\
  Z_{\perp}(\omega) =&\ Z_{\rm N}
  + Z_{{\rm FN}\perp} 
  + Z_{{\rm F}\perp}(\omega).
    \label{eq:Ztotal}
\end{align}
Using Eq.\ (\ref{eq:ZN1}) to calculate $\vmus(z \downarrow 0, \omega)$, Eqs.\ (\ref{eq:epolar}) and (\ref{eq:eperppolar}) for the unit vectors $\ve_{\parallel}$ and $\ve_{\perp}$, and Eqs. \eqref{eq:jcx1} and (\ref{eq:jcy1}), one can calculate the SMR corrections $\delta j_{\rm c}^x$ and $\delta j_{\rm c}^y$ to the current densities parallel and perpendicular to the applied electric field as
\begin{align}
  \delta \bar j^x_{\rm c}(\omega) =&\, \delta \sigma^{xx}(\omega) E(\omega),\nonumber \\
  \delta \bar j^y_{\rm c}(\omega) =&\, \delta \sigma^{xy}(\omega) E(\omega),
\end{align}
with
\begin{widetext}
\begin{align}
  \label{eq:jcx2}
  \delta \sigma^{xx}(\omega) =&\,
  \frac{\theta_{\rm SH}^2 \hbar \sigma_{\rm N}^2}{4 e^2 d_{\rm N}} Z_{\rm N}
  \left\{ 1
  - 
  m_y^2
  \frac{Z_{\rm N}}{Z_{\parallel} (\omega)} 
  - \frac{1 - m_y^2}{2} 
  \! \left[ \frac{Z_{\rm N}}{Z_{\perp}(\omega)}
  + \frac{Z_{\rm N}}{Z^*_{\perp}(-\omega)} \right] \!
  \right\}, \\
  \label{eq:jcy2}
  \delta \sigma^{xy}(\omega) =&\,
  \frac{\theta_{\rm SH}^2 \hbar \sigma_{\rm N}^2}{4 e^2 d_{\rm N}} Z_{\rm N}
  \left\{ 
   m_x m_y
   \! \left[\frac{Z_{\rm N}}{Z_{\parallel} (\omega)} 
  - \frac{Z_{\rm N}}{2 Z_{\perp}(\omega)}
  - \frac{Z_{\rm N}}{2 Z^*_{\perp}(-\omega)} \right] \!
  - \frac{i m_z}{2} 
  \! \left[\frac{Z_{\rm N}}{Z_{\perp}(\omega)}
  - \frac{Z_{\rm N}}{Z^*_{\perp}(-\omega)} \right] \!
  \right\}.
\end{align}
\end{widetext}

\noindent
These two equations are the central results of this article. They describe the longitudinal and transverse contributions to the SMR. The term ``1'' between the curly brackets in Eq.\ (\ref{eq:jcx2}), which does not depend on the interface properties, is the ISHE correction to the diagonal conductivity from a single non-magnetic insulating interface. The remaining terms, which depend on the properties of the F-N interface and on the magnetization direction $\vm_{\rm eq}$, describe the change of the conductivity corrections $\delta \sigma^{xx}$ and $\delta \sigma^{xy}$ due to the presence of the magnet F. The contribution to the off-diagonal conductivity $\delta \sigma^{xy}$ in Eq.\ (\ref{eq:jcy2}) proportional to $m_x m_y$ can be identified with a spin-Hall version of the planar Hall effect (PHE), which is symmetric under magnetization reversal. The terms proportional to $m_z$ correspond to a spin-Hall version of the anomalous Hall effect (AHE), which is antisymmetric under magnetization reversal. 

The interface to the vacuum at $z = d_{\rm N}$, which is not considered here, gives an additional correction to $\sigma^{xx}$. In the limit of large $d_{\rm N}$ this correction is given by the term ``1'' between the curly brackets in Eq.\ (\ref{eq:jcx2}).

\begin{figure}
\ctikzset{bipoles/length=.75cm}
\centering
\hspace{-0.425cm}
\begin{minipage}{.25\textwidth}
\begin{circuitikz}[european inductors]
\node[anchor=north west,inner sep=0] at (-0.25,-0.9) {\includegraphics[width=0.99\textwidth]{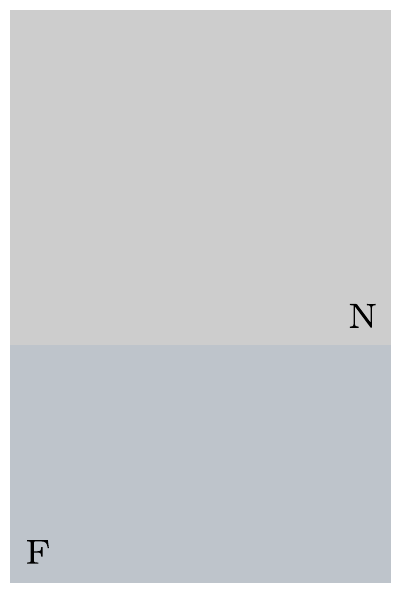}};
\draw[densely dashed] (4,0) 
to[short] (0,0)
to[short] (0,-1.5)
to[generic=$\sigma_{\rm N}$] (4,-1.5)
to[short] (4,0);
\fill [white] (2,0) circle (1.9ex);
\draw (2.25,0)
to[sV, l=$E$] (1.75,0);
\draw (3.5,-1.5)
to[tgeneric, i<^=$\delta j_{\rm c \perp}^x \frac{d_{\rm N}}{\hbar}$] (3.5,-4)
to[short] (0.5,-4)
to[tgeneric, i>^=$j_{\rm SH \perp}^z$] (0.5,-1.5);
\draw (2,-2.5)
to[generic=$Z_{{\rm N}}$, i<^=$j_{{\rm sN}}^z$] (2,-4);
\draw (2,-2.8)
node[ground, yscale=-1]{};
\draw (2,-4)
to[generic=$Z_{{\rm FN}\perp}$] (2,-5.5)
to[L=$Z_{{\rm F} \perp}$, i^<=$j_{{\rm s}\perp}^z(0)$] (2,-7.4);
\draw (2,-7.2)
node[ground]{};
\draw (0.5,-1.5) node[circ]{};
\draw (3.5,-1.5) node[circ]{};
\draw (2,-4) node[circ]{};
\draw (2,-4) node[anchor=north east]{$\mu_{{\rm s}\perp}(z\! \downarrow \!0)$};
\draw (2,-5.6) node[circ]{};
\draw (2,-5.6) node[anchor=east]{$- \hbar \omega m_{\perp}(0)$};
\draw (0.25,-2.75) node[anchor=south, rotate=90] {$\sigma_{{\rm SH}\perp}$}; 
\draw (3.75,-2.75) node[anchor=north, rotate=90] {$\sigma_{{\rm SH}\perp}$}; 
\end{circuitikz}
\end{minipage}%
\begin{minipage}{.25\textwidth}
\begin{circuitikz}[european inductors]
\node[anchor=north west,inner sep=0] at (-0.25,-0.9) {\includegraphics[width=0.99\textwidth]{./circuit_background_lighter_narrower.pdf}};
\draw[densely dashed] (4,0) 
to[short] (0,0)
to[short] (0,-1.5)
to[generic=$\sigma_{\rm N}$] (4,-1.5)
to[short] (4,0);
\fill [white] (2,0) circle (1.9ex);
\draw (2.25,0)
to[sV] (1.75,0);
\draw (3.5,-1.5)
to[tgeneric, i<^=$\delta j_{\rm c \parallel}^x \frac{d_{\rm N}}{\hbar}$] (3.5,-4)
to[short] (0.5,-4)
to[tgeneric, i>^=$j_{\rm SH \parallel}^z$] (0.5,-1.5);
\draw (2,-2.5)
to[generic=$Z_{{\rm N}}$, i<^=$j_{{\rm sN}}^z$] (2,-4);
\draw (2,-2.8)
node[ground, yscale=-1]{};
\draw (1.25,-4)
to[generic=$Z^{\rm e}_{{\rm FN}\parallel}$] (1.25,-5.5)
to[L=$Z^{\rm e}_{{\rm F} \parallel}$, i^<=$j_{{\rm se}\parallel}^z(0)$] (1.25,-7.4);
\draw (1.25,-7.2)
node[ground]{};
\draw (2.75,-4)
to[generic=$Z^{\rm m}_{{\rm FN}\parallel}$] (2.75,-5.5)
to[L=$Z^{\rm m}_{{\rm F} \parallel}$, i^<=$j_{{\rm sm}\parallel}^z(0)$] (2.75,-7.4);
\draw (2.75,-7.2)
node[ground]{};
\draw (0.5,-1.5) node[circ]{};
\draw (0.5,-1.5) node[anchor=south east]{$E$};
\draw (3.5,-1.5) node[circ]{};
\draw (1.25,-4) node[circ]{};
\draw (1.25,-4) node[anchor=north east]{$\mu_{{\rm s}\parallel}(z \!\downarrow\!0)$};
\draw (1.25,-5.6) node[circ]{};
\draw (1.25,-5.6) node[anchor=east]{$\mu_{{\rm s}\parallel}(z \!\uparrow \!0)$};
\draw (2.75,-5.6) node[circ]{};
\draw (2.75,-5.6) node[anchor=east]{$\mu_{\rm m}(0)$};
\draw (2,-4) node[circ]{};
\draw (2.75,-4) node[circ]{};
\draw (0.25,-2.75) node[anchor=south, rotate=90] {$\sigma_{{\rm SH}\parallel}$}; 
\draw (3.75,-2.75) node[anchor=north, rotate=90] {$\sigma_{{\rm SH}\parallel}$}; 
\end{circuitikz}
\end{minipage}%
\caption{Equivalent AC magnetoelectronic circuit diagrams for $\delta \sigma^{xx}$. The left circuit shows the correction from the transverse spin accumulation at the F-N interface; the right one shows the correction from the longitudinal spin accumulation. The dashed lines in the upper part indicate the transport of charge, while the solid lines in the lower part indicate the transport of spin angular momentum. Spin currents and spin accumulations are related via (spin versions of) Ohm's law and Kirchoff's circuit laws, see Eqs.\ \eqref{eq:jcx1}--\eqref{eq:ZFN1}. The impedances with labels $\sigma_{\rm SH}$ and arrows indicate the conversion between charge and spin currents due to the SHE and ISHE, see Eqs.\ (\ref{eq:jcx})--(\ref{eq:jsz}). The magnetization direction enters via the projection on longitudinal and transverse components, $\sigma_{{\rm SH}\parallel} = \theta_{\rm SH} \hbar \sigma_{\rm N} m_y/2e$ and $\sigma_{{\rm SH}\perp} = \theta_{\rm SH} \hbar \sigma_{\rm N} (1 - m_y^2)^{1/2}/2e$. The spin currents $j^z_{\rm SH \parallel}$ and $j^z_{\rm SH \perp}$ generated via the SHE are given by $j^z_{\rm SH \parallel} = \theta_{\text{SH}} \hbar \sigma_{\rm N} E \ve_{y} \cdot \ve_{\parallel}/2e = \sigma_{{\rm SH}\parallel} E$ and \mbox{$j^z_{\rm SH \perp} = \theta_{\text{SH}} \hbar \sigma_{\rm N} E \ve_{y} \cdot \ve_{\perp}/2e = \sigma_{{\rm SH}\perp} E$}. The fact that spin is not a conserved quantity --- neither in N due to flips of electron spins, nor in F due to Gilbert damping --- is reflected by the presence of ``spin sinks'' in N and F, represented as dissipation channels to the ``ground'' in the circuit diagram.}
\label{fig:equiv_circuit}
\end{figure}

The particular combination of impedances appearing in Eq.\ (\ref{eq:jcx2}) can be illustrated with the equivalent magnetoelectronic circuit diagrams in Fig.\ \ref{fig:equiv_circuit}, which shows the corrections to $\delta \sigma^{xx}$ from transverse and longitudinal spin currents and spin accumulations on the left and right, respectively. The currents and spin accumulations in the magnetoelectronic circuits of Fig.\ \ref{fig:equiv_circuit} are related via spin versions of Ohm's law and Kirchoff's circuit laws and thus fulfill Eqs.\ \eqref{eq:jcx1}--\eqref{eq:ZFN1} (cf. also Ref. \onlinecite{Zhu_2017}). 

In Sec.\ \ref{sec:impedances} we find that in the DC limit $\omega = 0$ one has $Z_{{\rm F}\perp}(0) = 0$. Additionally, for a typical thickness $d_{\rm F}$ of the F layer, $Z_{{\rm F} \parallel}^{\rm m} (0)$ is much larger than all other impedances, so that it is a very good approximation to not consider the contribution from the longitudinal magnon spin current. This is the approach taken in Refs.\ \onlinecite{Althammer_2013,Chen_2013,Chen_2016}. In this limit the contributions inversely proportional to $Z_{\perp}$ in Eqs.\ (\ref{eq:jcx2}) and (\ref{eq:jcy2}) agree with the DC theory of the SMR in an F$|$N bilayer for a ferromagnetic insulator of Refs.\ \onlinecite{Althammer_2013,Chen_2013,Chen_2016}. The contributions inversely proportional to $Z_{\parallel}$ is consistent with previous results for an F$|$N bilayer with a metallic ferromagnet.\cite{Kim_2016}

\section{Impedances} \label{sec:impedances}

As shown below, the spin impedances defined in Eqs. (\ref{eq:ZN1})--(\ref{eq:ZFN1}) characterize different physical processes that affect spin transport:
\begin{itemize}
\item
$Z_{\rm N}$ the relaxation of the diffusing spin accumulation in N (due to magnetic and non-magnetic impurities, phonons via spin-orbit coupling, and other spin-flip mechanisms); 
\item
$Z_{{\rm F} \parallel}^{\rm e}$ the same as $Z_{\rm N}$, but for a ferromagnetic metal F; 
\item
$Z_{{\rm FN} \parallel}^{\rm e}$ the spin current carried by electrons transported across an F-N interface; 
\item
$Z_{{\rm F} \perp}$ the transport and relaxation of spin currents by coherent magnons in F; 
\item 
$Z_{{\rm FN} \perp}$ the spin-transfer torque and spin pumping at an F-N interface; 
\item
$Z_{{\rm FN} \parallel}^{\rm m}$ the incoherent creation and annihilation of thermal magnons in F by spin-flip scattering of conduction electrons at an F-N interface; 
\item
and $Z_{{\rm F} \parallel}^{\rm m}$ the build-up and relaxation of a chemical potential for magnons in F (a ``magnon capacitance'' effect). 
\end{itemize}
In the following we describe the calculation of each of these spin impedances separately.

\subsection{Normal metal}
\label{subsec:normal_metal}

The continuity equation for spin currents and accumulations, including spin flips at a rate $1/\tau_{\rm sf,N}$ due to spin-orbit coupling and magnetic impurities, reads 
\begin{align}
  \frac{2}{\hbar}
  \frac{\partial }{\partial z} \vjs^{z} + \nu_{\rm N} \dot \vmus =
  - \frac{\nu_{\rm N} \vmus}{\tau_{\rm sf,N}},
\end{align}
where $\nu_{\rm N}$ is the density of states per spin direction. Combining this equation with the transport equation (\ref{eq:jsz}) and Fourier transforming to time, it follows that the spin accumulation $\vmus$ satisfies the spin-diffusion equation,
 \begin{equation}
  \frac{\partial^2}{\partial z^2} \vmus(z,\omega) = \frac{\vmus(z,\omega)}{\lambda^2_{\rm N}}.
  \label{eq:spindiffusion}
\end{equation} 
The spin-relaxation length $\lambda_{\rm N}$ is given by
\begin{equation}
  \lambda^2_{\rm N} = \frac{\sigma_{\rm N} \tau_{\rm sf,N}/2 e^2 \nu_{\rm N}}{1 - i \omega \tau_{\text{sf,N}}}. \label{eq:spin_diffusion_length_N}
\end{equation}
The frequency dependence of $\lambda_{\rm N}$ can be neglected up to frequencies in the THz regime, because spin-flip rates $1/\tau_{\rm sf,N}$ in heavy metals with significant spin-Hall angle $\theta_{\rm SH}$ are typically much higher due to the strong spin-orbit coupling. For example, in Pt one has $1/\tau_{\rm sf,N} \sim 10^2 \, \rm{THz}$, see Ref.\ \onlinecite{Jiao_2013}. 

For $d_{\rm N} \gg \lambda_{\rm N}$, the solution of Eq.\ (\ref{eq:spindiffusion}) decays exponentially away from the F-N interface at $z = 0$. Integration over $z$ immediately yields Eq. (\ref{eq:ZN1}), with
\begin{equation}
Z_{\rm N} = \frac{4 e^2 \lambda_{\rm N}}{\hbar \sigma_{\rm N}}. \label{eq:ZN}
\end{equation}
The impedance $Z_{\rm N}$ is the resistance (normalized to cross-sectional area) for a normal-metal slab of thickness $\lambda_{\rm N}$.

\subsection{Ferromagnet: longitudinal contribution from electrons}
\label{subsec:ferromagnet_long}

In a metallic ferromagnet, conduction electrons can carry a spin current parallel to the magnetization direction $\ve_{\parallel}$. The calculation of this spin current component proceeds in the same manner as the calculation of the spin current density in the normal metal. Hereto, we introduce (electro-)chemical potentials $\mu_{\uparrow,\downarrow}$ and charge current densities $j_{{\rm c}\uparrow,\downarrow}^z$ carried by electrons with spin $\uparrow, \downarrow$, where now the arrows ``$\uparrow$'' and ``$\downarrow$'' indicate spin polarization parallel or antiparallel to $\ve_{\parallel}$, respectively. From the transport equations $j_{{\rm c}\uparrow,\downarrow}^z = - (\sigma_{\uparrow,\downarrow}/e) \partial \mu_{\uparrow,\downarrow}/\partial z$ for a metallic ferromagnet we derive that the charge current density $j_{\rm c}^z$ and spin current density $j_{{\rm se}\parallel}^z$ are
\begin{align}
  j_{{\rm c}}^z =&\,
  - \frac{\sigma_{{\rm c}}}{e} \frac{\partial }{\partial z} \mu_{\rm c} - \frac{\sigma_{\rm s}}{2 e} \frac{\partial }{\partial z} \mu_{{\rm s}\parallel}, \\
  j_{{\rm se}\parallel}^z =&\, - \frac{\hbar \sigma_{{\rm c}}}{4 e^2} \frac{\partial }{\partial z} \mu_{{\rm s}\parallel} - \frac{\hbar \sigma_{\rm s}}{2 e^2} \frac{\partial }{\partial z} \mu_{\rm c},
\end{align}
where $\mu_{\rm c} = (1/2)(\mu_{\uparrow} + \mu_{\downarrow})$, $\sigma_{{\rm c}} = \sigma_{\uparrow} + \sigma_{\downarrow}$, and $\sigma_{\rm s} = \sigma_{\uparrow} - \sigma_{\downarrow}$. Since charge density fluctuations are strongly suppressed by the long-range Coulomb interactions, we require that $j_{{\rm c}}^z = 0$, so that
\begin{equation}
  j_{{\rm se}\parallel}^z = - \frac{\hbar \sigma_{\rm F}}{4 e^2}
  \frac{\partial }{\partial z} \mu_{{\rm s}\parallel}, \quad\quad
  \sigma_{\rm F} = \frac{\sigma_{{\rm c}}^2 - \sigma_{\rm s}^2}{\sigma_{{\rm c}}}.
  \label{eq:jsF}
\end{equation}

The continuity equations for charge and spin read
\begin{align}
\frac{1}{e} \frac{\partial }{\partial z} j_{{\rm c}}^z + 2  \nu_{{\rm c}} \dot \mu_{\rm c} + \frac{ \nu_{\rm s}}{2} 
    \dot \mu_{{\rm s}\parallel}
  =&\, 0, 
  \\
  \frac{2}{\hbar}
\frac{\partial }{\partial z} j_{{\rm se}\parallel}^z + \nu_{{\rm c}} \dot \mu_{{\rm s}\parallel} + \nu_{\rm s} \dot \mu_{\rm c}
  =&\, - \frac{\nu_{\rm F}}{\tau_{\rm sf,F}} 
    \mu_{{\rm s}\parallel},
\end{align}
where $\tau_{{\rm sf,F}}$ is the phenomenological spin-flip time in the ferromagnet; $\nu_{{\rm c}} = (\nu_{\uparrow} + \nu_{\downarrow})/2$ and $\nu_{\rm s} = \nu_{\uparrow} - \nu_{\downarrow}$ are densities of states, and
\begin{equation}
  \nu_{\rm F} = \frac{4 \nu_{\rm c}^2 - \nu_{\rm s}^2}{4 \nu_{\rm c}}.
\end{equation}  
Again, using the absence of charge currents in the $z$ direction we eliminate $\dot \mu_{\rm c}$ from these equations and find
\begin{align}
  \frac{2}{\hbar}
  \frac{\partial }{\partial z} j_{{\rm se}\parallel}^z
  + \nu_{\rm F} \dot \mu_{{\rm s}\parallel}
  =&\,  -\frac{\nu_{\rm F}}{\tau_{\rm sf,F}} \mu_{{\rm s}\parallel}.
  \label{eq:jscontinuityF}
\end{align}
Combining Eqs.\ (\ref{eq:jscontinuityF}) and (\ref{eq:jsF}) and performing a Fourier transform to frequency, we obtain the spin-diffusion equation
\begin{equation}
  \frac{\partial^2 }{\partial z^2} \mu_{{\rm s}\parallel}(z,\omega)
  = \frac{\mu_{{\rm s}\parallel}(z,\omega)}{\lambda_{\rm F}^2},
\end{equation}

\noindent
with the spin-relaxation length $\lambda_{\rm F}$ of F given by
\begin{equation}
  \lambda_{\rm F}^2 = \frac{\sigma_{\rm F} \tau_{\rm sf,F}/2 e^2 \nu_{\rm F}}{1 - i \omega \tau_{\rm sf, F}}.
  \label{eq:spindiffusionF}
\end{equation}
As in the case of the normal metal N, spin flip rates $1/\tau_{\rm sf,F}$ in metallic ferromagnets are assumed to be large, so that we may safely ignore the frequency dependence of $\lambda_{\rm F}$ for the frequencies of interest in the THz range and below.

Solving Eq.\ \eqref{eq:spindiffusionF} with the boundary condition $j_{{\rm se}\parallel}^z (-d_{\rm F}, \omega) = 0$, one finds
\begin{equation}
  \mu_{{\rm s}\parallel}(z,\omega) = \mu_{{\rm s}\parallel}(z \uparrow 0,\omega) \frac{\cosh[(z+d_{\rm F})/\lambda_{\rm F}]}{\cosh(d_{\rm F}/\lambda_{\rm F})}.
\end{equation}
\noindent
The spin current $j^{z}_{{\rm se}\parallel}$ at the F-N interface is
\begin{equation}
  j_{{\rm se}\parallel}^z(0,\omega) = - \frac{\hbar \sigma_{\rm F}}{4 e^2 \lambda_{\rm F}} \tanh \!\left( \frac{d_{\rm F}}{\lambda_{\rm F}} \right) \mu_{{\rm s}\parallel}(z \uparrow 0,\omega) .
  \label{eq:jsF1}
\end{equation}
Comparing with Eq.\ (\ref{eq:ZF1}) we conclude that 
\begin{equation}
  Z^{\rm e}_{{\rm F}\parallel} = \frac{4 e^2 \lambda_{\rm F}}{\hbar \sigma_{\rm F}} \coth \!\left( \frac{d_{\rm F}}{\lambda_{\rm F}} \right).
\end{equation}

\subsection{F-N interface: longitudinal contribution from electrons}
\label{subsec:interface_long_el}

The linearized charge and spin currents through the F-N interface at $z=0$ collinear with $\vm$ are given by the equations\cite{Tserkovnyak_2005}
\begin{align}
  j_{\rm c}^z(0,\omega) =&
  - \frac{e}{h} g_{\rm c} \Delta \mu_{\rm c}(\omega) 
  - \frac{e}{2 h} g_{\rm s} \Delta \mu_{{\rm s}\parallel}(\omega) , \\
  j_{{\rm se}\parallel}^z(0,\omega) =&\,
  - \frac{g_{\rm s}}{4 \pi}  \Delta \mu_{\rm c}(\omega) 
  - \frac{g_{\rm c}}{8 \pi}  \Delta \mu_{{\rm s}\parallel}(\omega) ,
  \label{eq:jsinterfaceparallel}
\end{align}
where $\Delta \mu_{\rm c}(\omega) = \mu_{\rm c}(z \downarrow 0,\omega) - \mu_{\rm c}(z \uparrow 0,\omega)$, $\Delta \mu_{{\rm s}\parallel}(\omega) = \mu_{{\rm s}\parallel}(z \downarrow 0,\omega) - \mu_{{\rm s}\parallel}(z \uparrow 0,\omega)$ are the drops of potential and spin accumulation over the F-N interface, $g_{\rm c} = g_{\uparrow\uparrow} + g_{\downarrow\downarrow}$ is the total dimensionless interface conductance per unit area, and $g_{\rm s} = g_{\uparrow\uparrow} - g_{\downarrow\downarrow}$. For a ferromagnetic insulator $g_{\uparrow\uparrow} = g_{\downarrow\downarrow} = 0$ and, hence, $g_{\rm c} = g_{\rm s} = 0$.

As before, we require that there be no charge current through the interface, which gives
\begin{equation}
  j_{{\rm se}\parallel}(0,\omega) = 
  - \frac{g_{\rm FN}}{8 \pi} 
  \Delta \mu_{{\rm s}\parallel}(\omega),
  \label{eq:j3interface}
\end{equation}
with
\begin{equation}
  g_{\rm FN} = \frac{g_{\rm c}^2 - g_{\rm s}^2}{g_{\rm c}}.
\end{equation}
Comparing Eq.\ (\ref{eq:j3interface}) with Eq.\ (\ref{eq:ZFN0}), we conclude that
\begin{equation}
  Z_{{\rm FN}\parallel}^{\rm e} = \frac{8 \pi}{g_{\rm FN}}.
\end{equation}
Since the reflection at the F-N interface is effectively instantaneous, the frequency dependence $Z^{\rm e}_{{\rm FN}\parallel}$ may be safely neglected for frequencies in the THz range and below.

\subsection{Ferromagnet: transverse contribution}
\label{subsec:llg}

For both insulating and metallic ferromagnets, to linear order in the applied fields, spin currents transverse to the magnetization direction $\ve_{\parallel}$ are carried by magnons. Up the THz frequency range, the resulting magnetization dynamics may be calculated from the Landau-Lifshitz-Gilbert equation. This approximation is valid as long as the frequency of the applied fields is far below the excitation threshold of optical and zone-boundary magnons, which, for YIG, holds for frequencies up to $\omega/2 \pi \approx 5\, \rm{THz}$.\cite{Barker_2016}

The Landau-Lifshitz-Gilbert equation reads\cite{Gilbert_1955,Landau_1980}
\begin{align}
  \label{eq:llg}
  \dot \vm =&\, \omega_0 \ve_{\parallel} \times \vm + \alpha \vm \times \dot \vm - \Dex \vm \times \frac{\partial^2 \vm}{\partial z^2}.
\end{align}
Here $\omega_0$ is the ferromagnetic-resonance frequency, which includes effects of static external magnetic fields, demagnetization field, and anisotropies; $\alpha$ is the bulk Gilbert damping coefficient, and $\Dex$ the spin stiffness. The spin current density from the magnetization dynamics is\cite{Barnes_2006}
\begin{equation}
  \vjs^z = - \frac{\Dex M_{\rm s}}{\gamma} \vm \times \frac{\partial \vm}{\partial z},
  \label{eq:jsllg}
\end{equation}
where $M_{\rm s}$ is the magnetic moment per unit volume and $\gamma = \mu_{\rm B} g/\hbar$ the gyromagnetic ratio. 

Inserting the parameterization (\ref{eq:m1}) and keeping terms to linear order in $m_{\perp}$ only, one gets the linearized Landau-Lifshitz-Gilbert equation
\begin{align}
  -\Dex
    \frac{\partial^2 }{\partial z^2} m_{\perp}(z,\omega)
  =&\,
  (\omega + i \alpha \omega - \omega_0) m_{\perp}(z,\omega).
\end{align}
With the boundary condition that the spin currents must vanish at the boundary of the F layer at $z=-d_{\rm F}$, the solution of this equation is
\begin{align}
  m_{\perp}(z,\omega) =&\, 
    \frac{\cos[K(\omega) (z + d_{\rm F})]}
      {\cos[K(\omega) d_{\rm F}]} 
    m_{\perp}(0,\omega),
\end{align}
where $K(\omega)$ is the solution of
\begin{equation}
  \omega_0 + \Dex K^2 = \omega(1 + i \alpha).
  \label{eq:k}
\end{equation}
For the transverse spin current through the F-N interface at $z=0$ we then find
\begin{align}
  j_{{\rm s}\perp}^z(0,\omega) =&\,
  - \frac{i \Dex M_{\rm s}}{\gamma}
  K(\omega) \tan[K(\omega) d_{\rm F}] m_{\perp}(0,\omega),
\end{align}
so that
\begin{equation}
  \label{eq:ZF2}
  Z_{{\rm F}\perp}(\omega) = i 
  \frac{\hbar \gamma \omega}{\Dex M_{\rm s} K(\omega)} 
  \cot[K(\omega) d_{\rm F}].
\end{equation}

To further analyze this expression, we separate real and imaginary parts of the complex wavenumber $K(\omega)$,
\begin{equation}
  K(\omega) = k(\omega) + i \kappa(\omega).
\end{equation}
For frequencies $\omega > \omega_0$ magnon modes exist in F. For small Gilbert damping $\alpha$, the complex wavenumber $K(\omega)$ is close to being real if $\omega > \omega_0$, with small imaginary part
\begin{equation}
  \kappa(\omega) \approx \frac{\alpha \omega}{v(\omega)},\ \ \omega \gtrsim \omega_0,
  \label{eq:kappapositive}
\end{equation}
where $v(\omega) = d\omega/dk$ is the magnon velocity. For $\omega < \omega_0$, which includes negative frequencies, the complex wavenumber $K(\omega)$ is close to being purely imaginary, with imaginary part
\begin{equation}
  \kappa(\omega) \approx \sqrt{\frac{\omega_0-\omega}{\Dex}},\ \
  \omega \lesssim \omega_0,
  \label{eq:kappanegative}
\end{equation}
reflecting the absence of magnon modes at these frequencies.

In the limit of large $d_{\rm F}$, such that $\kappa(\omega) d_{\rm F} \gg 1$, one has $\cot[k(\omega) d_{\rm F}] \to -i$, so that
\begin{equation}
  Z_{{\rm F}\perp}(\omega) \to Z_{{\rm F}\perp}^{\infty}(\omega) \equiv
  \frac{\hbar \gamma \omega}{\Dex M_{\rm s} K(\omega)}.
  \label{eq:Zinfinity}
\end{equation}
For $|\omega| \gg \omega_0$, the limiting impedance $Z_{{\rm F}\perp}^{\infty}(\omega)$ may be approximated as
\begin{equation}
  Z_{{\rm F}\perp}^{\infty}(\omega) \approx
  \frac{\hbar \gamma}{M_{\rm s}}\sqrt{\frac{|\omega|}{\Dex}} \times \left\{
  \begin{array}{ll} 1 & \mbox{if $\omega \gg \omega_0$}, \\
    -i & \mbox{if $\omega \ll -\omega_0$}. \end{array} \right.
  \label{eq:Zapprox}
\end{equation}
Upon going to smaller thicknesses $d_{\rm F}$ of the ferromagnetic layer, the approximation $\kappa(\omega) d_{\rm F} \gg 1$ first breaks down for frequencies $\omega \gtrsim \omega_0$, because the imaginary part $\kappa(\omega)$ is smallest in that case, see Eqs.\ (\ref{eq:kappapositive}) and (\ref{eq:kappanegative}).
To analyze the impedance $Z_{{\rm F}\perp}(\omega)$ in the regime $\kappa(\omega) d_{\rm F} \ll 1$ for frequencies $\omega \gtrsim \omega_0$, we note that 
\begin{align}
  \frac{Z_{{\rm F}\perp}(\omega)}{Z_{{\rm F}\perp}^{\infty}(\omega)} \approx&\,
  i \cot[k(\omega) d_{\rm F}] \nonumber \\ &\, \mbox{}
  + \sum_{n} \frac{\kappa(\omega) d_{\rm F}}{(k(\omega) d_{\rm F} - n \pi)^2
    + \kappa(\omega)^2 d_{\rm F}^2},
    \label{eq:resonance_pattern}
\end{align}
where the summation is over the integers $n$. The real part of $Z_{{\rm F}\perp}(\omega)/Z_{{\rm F}\perp}^{\infty}(\omega)$ exhibits a resonance structure with resonance spacing $\Delta \omega \approx \pi v/d_{\rm F}$ and a Lorentzian line shape with height $1/\kappa(\omega) d_{\rm F} \approx v/\alpha \omega d_{\rm F}$ and full width at half maximum $\approx 2 \alpha \omega$. It averages to one if averaged over a frequency window of width much larger than the resonance spacing $\Delta \omega$, but smaller than $\omega-\omega_0$. The imaginary part of $Z_{{\rm F}\perp}(\omega)/Z_{{\rm F}\perp}^{\infty}(\omega)$ shows large oscillations with period $\Delta \omega$ that average to zero if averaged over frequency.

The magnetic field of the driving field and the (Oersted) field of the alternating charge current give an additional correction to the transverse impedance discussed here and, hence, to the SMR. Since these fields are spatially uniform, they mainly couple to the uniform precession mode. Their effect on the SMR is strongest for frequencies in the vicinity of the ferromagnetic-resonance frequency, but it is negligible for other frequencies. We refer to App.\ \ref{sec:magnonic_spin_F} for a more detailed discussion.

\subsection{F-N interface: transverse contribution}
\label{subsec:interface}

The transverse spin current across the F-N interface couples to the magnetization dynamics via the spin transfer torque and spin pumping,\cite{Tserkovnyak_2005} 
\begin{align}
  j_{{\rm s}\perp}^z(0,\omega) =&\,
    - \frac{g_{\uparrow\downarrow}}{4 \pi}
    [\mu_{{\rm s}\perp}(z \downarrow 0,\omega) + \hbar \omega m_{\perp}(0,\omega)],
  \label{eq:jsinterface}
\end{align}
where $g_{\uparrow\downarrow}$ is the dimensionless spin-mixing conductance per unit area. Using the Landauer-B\"uttiker approach, $g_{\uparrow\downarrow}$ is defined as\cite{Brataas_2000}
\begin{align}
  g_{\sigma\sigma'} =&\, \frac{1}{A}
    \mbox{tr}\, (1- r_{\sigma} r_{\sigma'}^{\dagger}), \ \
  \sigma,\sigma'=\uparrow,\downarrow,
\end{align}
with $A$ the total area of the F-N interface and $r_{\uparrow}$ and $r_{\downarrow}$ the reflection matrix for majority and minority electrons, respectively. Comparing Eqs.\ (\ref{eq:jsinterface}) with Eq.\ (\ref{eq:ZFN1}), we obtain
\begin{equation}
  Z_{{\rm FN}\perp} = \frac{4 \pi}{g_{\uparrow\downarrow}}.
\end{equation}

\subsection{F-N interface: longitudinal magnon contribution}
\label{subsec:ferromagnet_long_mag}

As described in Subsec. \ref{subsec:llg}, a spin wave at frequency $\Omega/2 \pi$ carries an alternating spin current $j^z_{{\rm s}\perp}$ with spin polarization perpendicular to the magnetization direction $\ve_{\parallel}$. The magnitude of this transverse spin current is proportional to the amplitude $m_{\perp}$ of the spin wave, see Eq.\ (\ref{eq:jsllg}). Additionally, a spin wave carries a non-oscillating ({\em i.e.}, steady-state) spin current $j^z_{{\rm sm}\parallel}$ with spin polarization collinear with the magnetization direction. The magnitude of this longitudinal spin current is proportional to the difference of the {\em densities} of magnons moving in the positive and negative $z$ direction, which is quadratic in $m_{\perp}$. The net longitudinal magnon current across the F-N interface is nonzero if the F and N layers are brought out of equilibrium, which occurs, {\em e.g.}, if the spin accumulation $\mu_{{\rm s}\parallel}(z \downarrow 0)$ in N is nonzero.\cite{Flipse_2014} In Refs.\ \onlinecite{Bender_2015,Cornelissen_2016} the longitudinal spin current is calculated to leading order in the spin-mixing conductance $g_{\uparrow\downarrow}$ of the F-N interface (see also Refs.\ \onlinecite{Schmidt_2018,Reiss_2021})
\begin{align} \label{eq:jsm}
  j_{{\rm sm}\parallel}^z(0) =&\, - \frac{\hbar \gamma}{\pi M_{\rm s}} \mbox{Re} (g_{\uparrow\downarrow})
  \int d\Omega \, \nu_{\rm m}(\Omega) \Omega \left( -\frac{df^0}{d\Omega} \right)
  \nonumber \\  &\, \mbox{} \times
  [\mu_{{\rm s}\parallel}(z \downarrow 0) - \mu_{{\rm m}}(0)],
\end{align}
where $f^0(\Omega) = 1/(e^{\hbar \Omega/k_{\rm B} T}-1)$ is the Planck function, \textit{i. e.}, a Bose--Einstein distribution with zero chemical potential, and $\nu_{\rm m}(\Omega) = (\Omega-\omega_0)^{1/2}/(4 \pi^2 \Dex^{3/2})$ the magnon density of states. Equation (\ref{eq:jsm}) assumes that the temperatures on both sides of the F-N interface are the same.

Since in our system the spin accumulation $\mu_{{\rm s}\parallel}(z \downarrow 0)$ oscillates at frequency $\omega/2\pi$ due to the applied alternating electric field and the SHE, the associated longitudinal magnon current $j_{{\rm sm}\parallel}^z$ in F oscillates at the same frequency. For driving frequencies $\omega \lesssim k_{\rm B} T/\hbar$, this alternating longitudinal magnon spin current can be obtained from the steady-state result (\ref{eq:jsm}).
For the spin impedance $Z_{{\rm FN}\parallel}^{\rm m}$ it follows that
\begin{align}
  \frac{1}{Z_{{\rm FN}\parallel}^{\rm m}} =&\, \frac{\hbar \gamma}{\pi M_{\rm s}} \mbox{Re} (g_{\uparrow\downarrow})
  \int d\Omega \, \nu_{\rm m}(\Omega) \Omega \left( -\frac{df^0}{d\Omega} \right).
  \label{eq:ZFNparallel}
\end{align}
Corrections to Eq.\ (\ref{eq:ZFNparallel}) from the breakdown of the adiabatic approximation will become relevant at frequencies $\omega \gtrsim k_{\rm B} T/\hbar$. At room temperature this is at \mbox{$\omega/2 \pi \gtrsim 6\, {\rm THz}$}.

Eq. (\ref{eq:ZFNparallel}) may be further simplified in the limit $\hbar \omega_0 \ll k_{\rm B} T$, which is applicable at room temperature. In this limit one finds\cite{Cornelissen_2016}
\begin{equation}
  \frac{1}{Z_{{\rm FN}\parallel}^{\rm m}} \approx
  \frac{3 \hbar \gamma\zeta(3/2)}{16 M_{\rm s} \pi^{5/2} }k_{\rm T}^3 \, \mbox{Re}\, g_{\uparrow\downarrow},
  \label{eq:ZFNparallelfinal}
\end{equation}
where $k_{\rm T} = \sqrt{k_{\rm B} T/\hbar \Dex}$ is the thermal magnon wavenumber and $\zeta(3/2) \approx 2.61$.

\subsection{Ferromagnet: longitudinal magnon contribution}
\label{subsec:ferromagnet_long_mag_diffusive}

If the thickness $d_{\rm F}$ of the F layer is so small and the magnon lifetime in F so long that the excess magnons excited at the F-N interface cannot be transported away from the interface and dissipated in the bulk F efficiently enough, a finite magnon chemical potential $\mu_{{\rm m}}(\omega)$ in F builds up --- the magnet acts like a ``magnon capacitor''. We here describe this effect in the limit of small $d_{\rm F}$, assuming that relaxation processes conserving the magnon number are fast enough that the magnon distribution can be characterized by a uniform magnon chemical potential across the F layer. (The opposite limit, in which magnons propagate ballistically in F and do not relax, is discussed in App. \ref{subsec:ferromagnet_long_mag_ballistic}.)

Balancing the influx of magnons through the F-N interface and the decay of excess magnons with lifetime $\tau(\Omega) = 1/2 \alpha \Omega$ due to the phenomenological Gilbert damping, we find, to linear order in $\mu_{{\rm m}}$,
\begin{align}
   j^z_{{\rm sm}\parallel} (0) =& - d_{\rm F} 
  \int_{\omega_0}^{\infty} d\Omega \, \nu_{\rm m}(\Omega) \!\left( - \frac{d f^0(\Omega)}{d\Omega} \right) \!
    \left[ \dot \mu_{{\rm m}} - \frac{\mu_{{\rm m}}}{\tau(\Omega)} \right] \!.
\end{align}
Fourier transforming and comparing to Eq.\ \eqref{eq:ZFm2} results in an impedance of the form
\begin{align}
  Z_{{\rm F} \parallel}^{\rm m} (\omega) = \frac{1}{C_{\rm m}(-i \omega + 1/\tau_{\rm m})}.
\label{eq:zFparallel_m}
\end{align}
In the limit $\hbar \omega_0 \ll k_{\rm B} T$, the expressions for the ``magnon capacitance'' $C_{\rm m}$ per unit area and the effective magnon life time $\tau_{\rm m}$ are
\begin{align}
  C_{\rm m} =&\, \frac{d_{\rm F}}{8 \pi \sqrt{\omega_0 \Dex}}
  k_{\rm T}^2,
  \label{eq:Cm} \\
  \tau_{\rm m} =&\, \frac{\sqrt{\pi}}{3 \zeta(3/2) \alpha k_{\rm T} \sqrt{\omega_0 \Dex}},
\label{eq:rFparallel_m}
\end{align}
where the thermal magnon wavenumber $k_{\rm T}$ is defined below Eq.\ (\ref{eq:ZFNparallelfinal}). The effective magnon lifetime $\tau_{\rm m}$ is significant in the low-frequency regime $\omega \tau_{\rm m} \lesssim 1$ only and may be effectively set to infinity for frequencies in the THz regime.

\subsection{Numerical estimates for the impedances}
\label{subsec:impedances_numerical_estimates}

To obtain an understanding of the order of magnitude of the spin impedances $Z_{\rm N}$, $Z^{\rm e}_{{\rm F}\parallel}$, $Z^{\rm m}_{{\rm F}\parallel}$, $Z_{{\rm F}\perp}$, $Z_{{\rm FN}\parallel}^{\rm e}$, $Z_{{\rm FN}\parallel}^{\rm m}$, and $Z_{{\rm FN}\perp}$, we calculate numerical values using typical parameters for an F$|$N bilayer consisting of the ferromagnetic insulator YIG and the normal metal Pt, as well as for a bilayer consisting of the ferromagnetic metal Fe and the normal metal Au. Numerical values for the relevant material and device parameters are collected in Tab.\ \ref{tab:estimates_parameters}, together with experimental references. (We note, however, that there is a large variation in literature values for the spin-Hall angle\cite{Hahn_2013,Hahn_2013_2,Althammer_2013,Nakayama_2013,Wang_2014,Vlietstra_2013,Qiu_2013,Mosendz_2010,Hoffmann_2013,Czeschka_2011,Liu_2012_2,Morota_2011} $\theta_{\rm SH}$, the spin-relaxation lengths\cite{Hahn_2013_2,Althammer_2013,Nakayama_2013,Wang_2014,Vlietstra_2013,Qiu_2013,Castel_2012,Bass_2007} $\lambda_{\rm N, F}$, and the interface conductances\cite{Hahn_2013,Hahn_2013_2,Althammer_2013,Nakayama_2013,Wang_2014,Vlietstra_2013,Qiu_2013} $g_{\sigma\sigma'}$. Different values for these quantities lead to different quantitative predictions, but do not affect our qualitative conclusions.)

\begin{table}
\centering
\begin{tabular*}{\columnwidth}{@{\extracolsep{\fill}}lll}
\hline
\hline
 & material and device parameters & ref. \\
\hline
Pt & $\sigma_{\rm N} = 9 \cdot 10^6 \, \Omega^{-1} \rm{m}^{-1}$, & \cite{Corti_1984}\\
& $\theta_{\rm SH} = 0.1 $, $\lambda_{\rm N} = 2 \cdot 10^{-9}\, {\rm m}$,  & \cite{Althammer_2013,Weiler_2013,Lotze_2014}\\
& $d_{\rm N} = 4 \cdot 10^{-9}\, {\rm m}$ & \cite{Lotze_2014}\\\hline
Au & $\sigma_{\rm N} = 4 \cdot 10^7 \, \Omega^{-1} \rm{m}^{-1}$, &\cite{Matula_1979}\\
& $\theta_{\rm SH} = 0.08$, $\lambda_{\rm N} = 6 \cdot 10^{-8} \, \rm{m}$, & \cite{Wang_2014}\\
& $d_{\rm N} = 6 \cdot 10^{-8} \, \rm{m}$ & \cite{Urban_2001} \\ \hline
YIG & $\omega_0/2 \pi = 8 \cdot \rm{10^{9} \, Hz}$, $\alpha = 2 \cdot  10^{-4}$, & \cite{Hahn_2013}\\
& $\Dex = 8 \cdot 10^{-6} \, \rm{m}^2 \, \rm{s}^{-1}$, $M_{\rm s} = 1 \cdot 10^5 \, \rm{A \, m}^{-1}$, & \cite{Weiler_2013}\\
& $d_{\rm F} = 6 \cdot 10^{-8}\, {\rm m}$ & \cite{Lotze_2014} \\ \hline
Fe & $\omega_0/2 \pi = 8 \cdot \rm{10^{9} \, Hz}$, $\alpha =  5 \cdot  10^{-3}$, & \footnote{The resonance frequency $\omega_0$ depends strongly on the external magnetic field applied in experiments. To make a comparison possible, we assume a magnetic field which results in the same resonance frequency as for the YIG$|$Pt bilayers of Ref.\ \onlinecite{Hahn_2013}.}$^,$\cite{Urban_2001}\\
& $\Dex =  4 \cdot 10^{-6} \, \rm{m}^2 \, \rm{s}^{-1}$, $M_{\rm s} = 2 \cdot 10^6 \, \rm{A \, m}^{-1}$,  & \cite{Kittel_2005}\\
& $\sigma_{\rm F} =  1 \cdot 10^7 \, \Omega^{-1} \rm{m}^{-1}$, & \cite{Ho_1983}\\
& $\lambda_{\rm F} = 9 \cdot 10^{-9} \, \rm{m}$, & \cite{Bass_2007}\\
& $d_{\rm F} = 2 \cdot 10^{-8} \, \rm{m}$ & 
\cite{Urban_2001} \\ \hline
YIG$|$Pt & $(e^2/h) \textrm{Re} \, g_{\uparrow \downarrow} = 6 \cdot 10^{13} \, \Omega^{-1} \rm{m}^{-2} $, & \cite{Hahn_2013,Qiu_2013}\\
YIG$|$Pt & $(e^2/h) \textrm{Im} \, g_{\uparrow \downarrow} = 0.3 \cdot 10^{13} \, \Omega^{-1} \rm{m}^{-2} $ & \footnote{According to Refs. \onlinecite{Jia_2011,Althammer_2013}, $\textrm{Im} \, g_{\uparrow \downarrow}/\textrm{Re} \, g_{\uparrow \downarrow} \approx 0.05$, which is also used here to estimate $\textrm{Im} \, g_{\uparrow \downarrow}$.} \\
\hline
Fe$|$Au  & $(e^2/h) \textrm{Re} \, g_{\uparrow \downarrow} = 1 \cdot 10^{14} \, \Omega^{-1} \rm{m}^{-2} $ & \cite{Wang_2014}\\
(clean) & $(e^2/h) \textrm{Im} \,  g_{\uparrow \downarrow} = 0.05  \cdot 10^{14} \, \Omega^{-1} \rm{m}^{-2} $, & \cite{Zwierzycki_2005}\\
 & $(e^2/h) g_{\uparrow \uparrow} = 4 \cdot 10^{14} \, \Omega^{-1} \rm{m}^{-2}$, & \cite{Zwierzycki_2005}\\
& $(e^2/h) g_{\downarrow \downarrow} = 0.8 \cdot 10^{14} \, \Omega^{-1} \rm{m}^{-2}$ & \cite{Zwierzycki_2005}\\
\hline
\hline
\end{tabular*}\medskip

\caption{Typical values for the relevant material and device parameters of the F$|$N bilayers considered in this article. The last column states the references used for our estimates.}
\label{tab:estimates_parameters}
\end{table}

\begin{figure}
\centering
\includegraphics[width=0.5\textwidth]{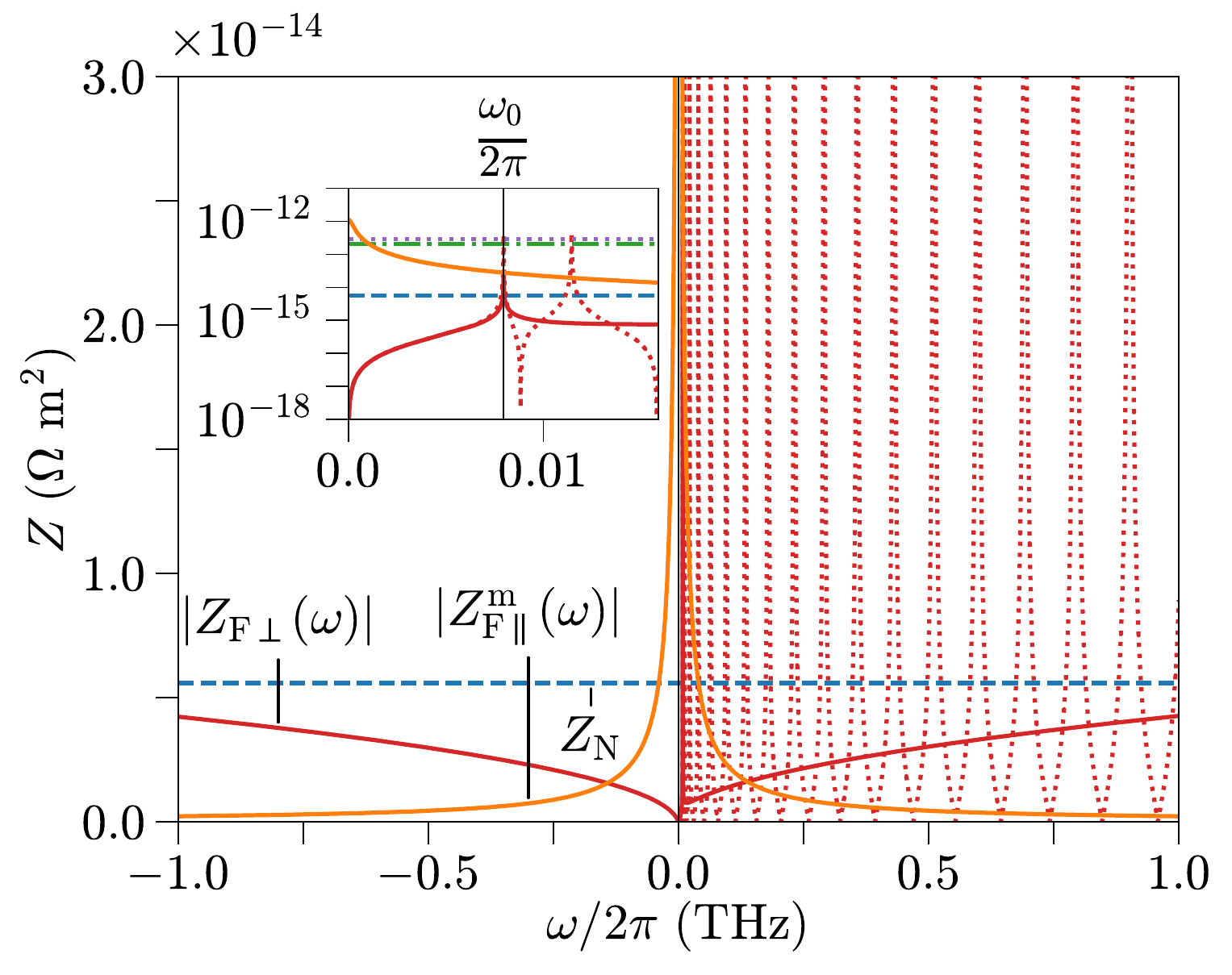}
\caption{Numerical estimates for the spin impedances of a YIG$|$Pt bilayer  
as a function of the frequency $\omega/2 \pi$. For the impedance $Z_{\rm N}$ the large-$d_{\rm N}$ limit is shown; for $|Z_{{\rm F}\perp}(\omega)|$ both the large-$d_{\rm F}$ limit (solid red line) and the finite-$d_{\rm F}$ case (dotted red line) are shown, as discussed in the main text. Parameter values are taken from Tab.\ \ref{tab:estimates_parameters}. The inset shows the same impedances on a logarithmic scale in the GHz frequency regime. The horizontal lines indicate $Z_{\rm N}$ (blue dashs, main panel and inset), $Z_{{\rm FN}\perp}$ (green dot-dashs, inset), and $Z_{{\rm FN}\parallel}^{\rm m}$ (light-purple dots, inset). \label{fig:YIG|Pt_impedances}}
\end{figure}

Estimates for the frequency-independent impedances as well as the ``magnon capacitance'' $C_{\rm m}$ and the effective magnon lifetime $\tau_{\rm m}$ obtained this way can be found in Tab.\ \ref{tab:estimates_impedances}. Since the spin-relaxation lengths $\lambda_{\rm N}$ and $\lambda_{\rm F}$ are smaller than typical layer thicknesses $d_{\rm N}$ and $d_{\rm F}$ used in experiments, we list $Z_{\rm N}$ and $Z_{{\rm F}\parallel}^{\rm e}$ for the limit of large $d_{\rm N}$ and $d_{\rm F}$, respectively. For the impedances $Z_{{\rm F}\perp}$ and $Z_{{\rm F}\parallel}^{\rm m}$, as well as $C_{\rm m}$, we take values for $d_{\rm F}$ typical for recent experiments, see Tab.\ \ref{tab:estimates_parameters}. Fig. \ref{fig:YIG|Pt_impedances} shows the frequency-dependent spin impedances $Z_{{\rm F}\perp}(\omega)$ and $Z_{{\rm F}\parallel}^{\rm m}(\omega)$ for a YIG$|$Pt bilayer. For a comparison of numerical values, the transverse spin impedance $Z_{{\rm F}\perp}^{\infty}(\omega)$ at $\omega/2 \pi = 1\, {\rm THz}$ and the longitudinal spin impedance $Z^{\rm m}_{{\rm F}\parallel}(\omega)$ for $\omega/2 \pi = 1\, {\rm GHz}$ are also included in Tab.\ \ref{tab:estimates_impedances}.
The impedance $Z_{{\rm F}\perp}(\omega)$ for a Fe$|$Au bilayer has a similar frequency dependence as in Fig.\ \ref{fig:YIG|Pt_impedances}, but a value that is a factor $\sim 10$ smaller (not shown). The smallness of $Z_{{\rm F}\perp}(\omega)$ in comparison to the transverse interface impedance $Z_{{\rm FN}\perp}$ for YIG$|$Pt and Fe$|$Au bilayers means that in both YIG and Fe spin angular momentum is efficiently transported away from the F-N interface for frequencies well into the THz regime.

For both YIG$|$Pt and Fe$|$Au, the longitudinal and transverse interface impedances are of comparable magnitude at $T = 300\,{\rm K}$. Since $|Z_{{\rm F}\perp}(\omega)|$ is typically much smaller than $|Z_{{\rm FN}\perp}|$, except in the immediate vicinity of resonances, the transverse spin current through the interface is dominated by the interfacial impedance. The same applies to the longitudinal spin current carried by conduction electrons if F is a metallic ferromagnet. The situation is different for the longitudinal magnonic spin current which depends strongly on frequency. In the zero-frequency limit, the longitudinal spin impedance $Z_{{\rm F}\parallel}^{\rm m}$ is much larger than the corresponding interfacial spin impedance, so that the longitudinal spin current carried by magnons is strongly suppressed. At high frequencies, $Z_{{\rm F}\parallel}^{\rm m}$ becomes small, and magnons can carry a sizeable longitudinal spin current. The crossover between these two regimes depends on the thickness $d_{\rm F}$ of the F layer and the ferromagnetic-resonance frequency $\omega_0$. For the parameters listed in Tab.\ \ref{tab:estimates_parameters}, the longitudinal magnon current sets in at $\omega/2 \pi \gtrsim 1\, {\rm GHz}$ for YIG$|$Pt. 

\begin{table}
  \centering
\begin{tabular*}{\columnwidth}{@{\extracolsep{\fill}}lll}
\hline
\hline
F$|$N & YIG$|$Pt & Fe$|$Au \\
\hline
$(h/e^2) Z_{\rm N}$ & $0.0055$ & $0.038$ \\
$(h/e^2) Z_{{\rm FN}\perp}$ & $0.21-0.01 i$ & $0.13 - 0.006 i$ \\
$(h/e^2) Z_{{\rm FN}\parallel}^{\rm m}$ & $0.30$ & $1.3$ \\
$(h/e^2) Z_{{\rm FN}\parallel}^{\rm e}$ & - & $0.096$ \\
$(h/e^2) Z^{\rm e}_{{\rm F}\parallel}$ & - & $0.023$ \\
$(e^2/h) C_{{\rm F}\parallel}^{\rm m}$ & $0.72$ & $0.68$ \\
$\tau_{\rm m}$ & $0.81$ & $0.032$ \\
$(h/e^2) Z_{{\rm F}\parallel}^{\rm m} (2 \pi \, \text{GHz})$ &
  $0.043+0.21i$ & $0.045 + 0.009i$ \\
$(h/e^2) Z_{{\rm F}\perp}^{\infty}(2 \pi\, \text{THz})$ &
$0.0043 - 4.3 \cdot 10^{-7} i$ & $0.0003 - 7.6 \cdot 10^{-7} i$ \\
\hline
\hline
\end{tabular*}\medskip

\caption{Estimates for the impedances of the F$|$N bilayers considered in this article. Parameter values are taken from Tab.\ \ref{tab:estimates_parameters}. All impedances are given in $10^{-12} \Omega\, {\rm m}^2$; the ``magnon capacitance'' $C_{\rm m}$ is in $10^{3}\, {\rm Fm}^{-2}$ and the effective magnon lifetime in $10^{-9} {\rm s}$. The interface impedance $Z_{{\rm FN}\parallel}^{\rm m}$, $C_{\rm m}$, and $\tau_{\rm m}$ are evaluated at $T = 300\, {\rm K}$. The estimates for $Z_{{\rm F}\parallel}^{\rm m}$ and $Z^{\infty}_{{\rm F}\perp}$ are at $\omega/2 \pi = 1\, {\rm GHz}$ and $\omega/2\pi = 1\, {\rm THz}$, respectively. The full frequency dependence of $Z_{{\rm F}\perp}(\omega)$ and $Z_{{\rm F}\parallel}^{\rm m}(\omega)$ is shown in Fig.\ \ref{fig:YIG|Pt_impedances} for YIG$|$Pt. }
\label{tab:estimates_impedances}
\end{table}

Taking typical values for $\omega_0$, $\alpha$, and $\Dex$ from Tab.\ \ref{tab:estimates_parameters}, one finds that at a frequency $\omega/2 \pi \sim 1$ THz the asymptotic large-$d_{\rm F}$ regime $\kappa(\omega) d_{\rm F} \gg 1$ for the transverse spin impedance $Z_{{\rm F}\perp}(\omega)$ sets in at $d_{\rm F} \gtrsim 1 \cdot 10^{-5}\, {\rm m}$ for YIG and $d_{\rm F} \gtrsim 3 \cdot 10^{-7}\, {\rm m}$ for Fe. (The difference is caused by the smallness of the Gilbert damping in YIG.) Both values are larger than the typical thickness $d_{\rm F} \sim 10^{-8}$m of thin magnetic films used in experiments, see Tab. \ref{tab:estimates_parameters}.
At these smaller thicknesses, $Z_{{\rm F}\perp}(\omega)$ develops a resonance structure, which may be measurable in experiments, if variations in $d_{\rm F}$ and frequency resolution are small enough. For example, at a thickness $d_{\rm F} \sim 6 \cdot 10^{-8}$ m of a YIG layer, the resonance spacing $\Delta \omega/2 \pi$ at frequency $\omega/2 \pi \sim 1$ THz is $\Delta \omega/2 \pi \sim 1 \cdot 10^{-1}$ THz, and for \mbox{$d_{\rm F} \sim 2 \cdot 10^{-8}$\, m} of an Fe layer $\Delta \omega/2 \pi \sim 2 \cdot 10^{-1}$ THz, whereas the typical frequency resolution of THz time-domain experiments is larger than $100$\,GHz, owing to the $<10$\,ps wide time window that is sampled.\cite{Jin_2015}

\section{Numerical estimates of the SMR}
\label{sec:experiments}

The conductivity corrections $\delta \sigma^{xx}$ and $\delta \sigma^{xy}$ depend on the magnetization direction $\vm_{\rm eq} = \ve_{\parallel}$, see Eq.\ (\ref{eq:epolar}). We characterize the conductivity corrections $\delta \sigma^{xx}(\ve_{\parallel})$ and $\delta \sigma^{xy}(\ve_{\parallel})$ corresponding to the spin-Hall magnetoresistance (SMR), planar Hall effect (PHE), and anomalous Hall effect (AHE) using the three complex dimensionless quantities
\begin{align}
  \Delta_{\rm SMR} =&\, [{\delta \sigma^{xx}(\ve_y) - \delta \sigma^{xx}(\ve_x)}]/{\sigma_{\rm N}}, \nonumber \\
  \Delta_{\rm PHE} =&\, -{\delta \sigma^{xy}(\ve_{xy})}/{\sigma_{\rm N}}, \nonumber \\
  \Delta_{\rm AHE} =&\, -{\delta \sigma^{xy}(\ve_{z})}/{\sigma_{\rm N}}, \label{eq:DeltaDef}
\end{align}
where we abbreviated $\ve_{xy} = (\ve_x + \ve_y)/\sqrt{2}$.
The planar Hall effect and spin-Hall magnetoresistance characteristics are related, see Eqs.\ (\ref{eq:jcx2}) and (\ref{eq:jcy2}),
\begin{equation}
  \Delta_{\rm PHE} = \frac{1}{2} \Delta_{\rm SMR}. \label{eq:DeltaPHE}
\end{equation}
Experimentally, the real part of $\Delta_{\rm SMR}$ is a magnetization direction-dependent correction to the magnitude of the current density $\bar j_{\rm c}^x(\omega)$ averaged over the thickness $d_{\rm N}$ of the N layer, whereas the imaginary part of $\Delta_{\rm SMR}$ is the magnetization direction-dependent part of the phase shift between $\bar j_{\rm c}^x(\omega)$ and the applied electric field $E(\omega)$. The modulus and phase of the complex coefficients $\Delta_{\rm PHE}$ and $\Delta_{\rm AHE}$ describe magnitude and phase of the transverse current $\bar j_{\rm c}^y(\omega)$.

\begin{figure}
\centering
\includegraphics[width=0.5\textwidth]{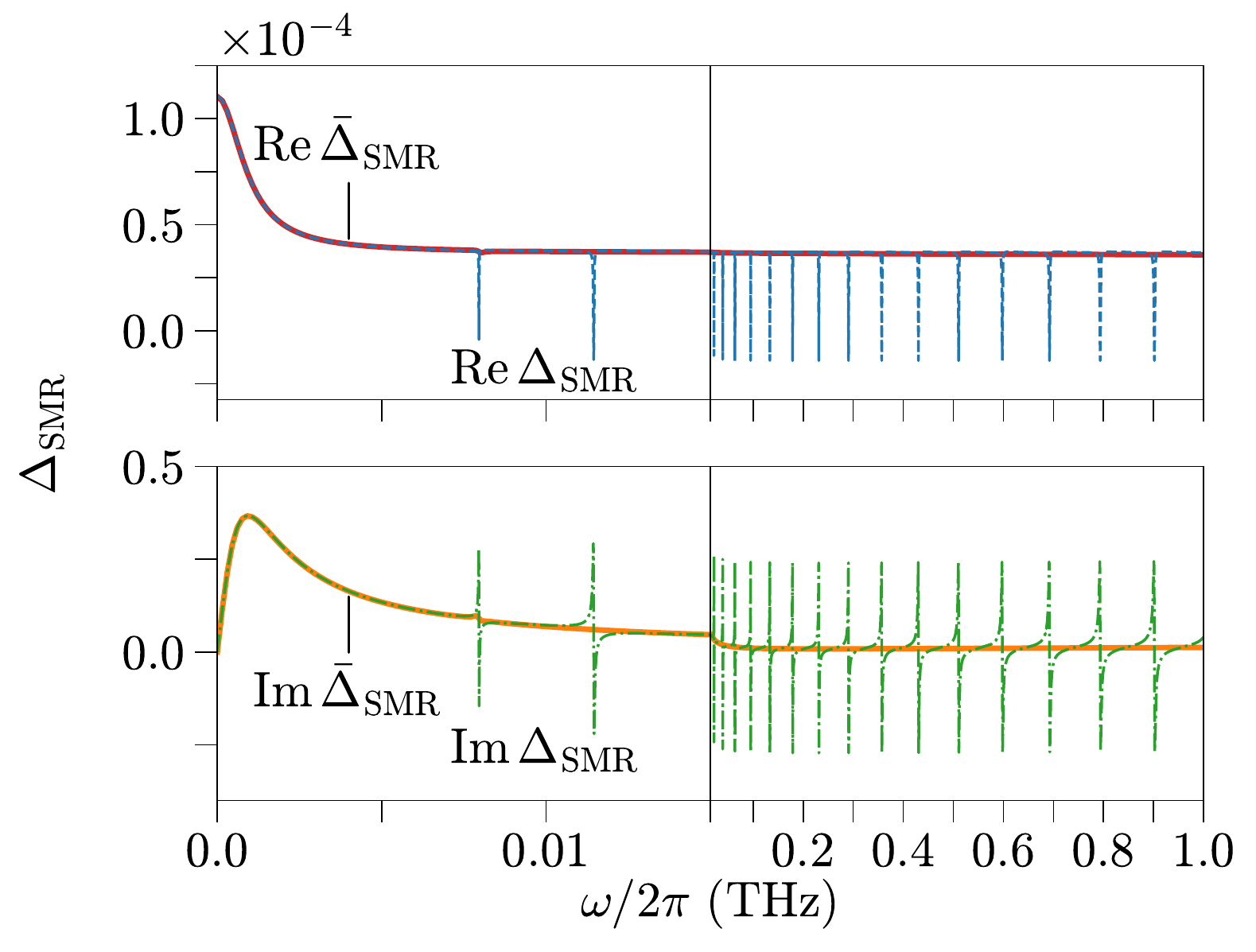}

\includegraphics[width=0.5\textwidth]{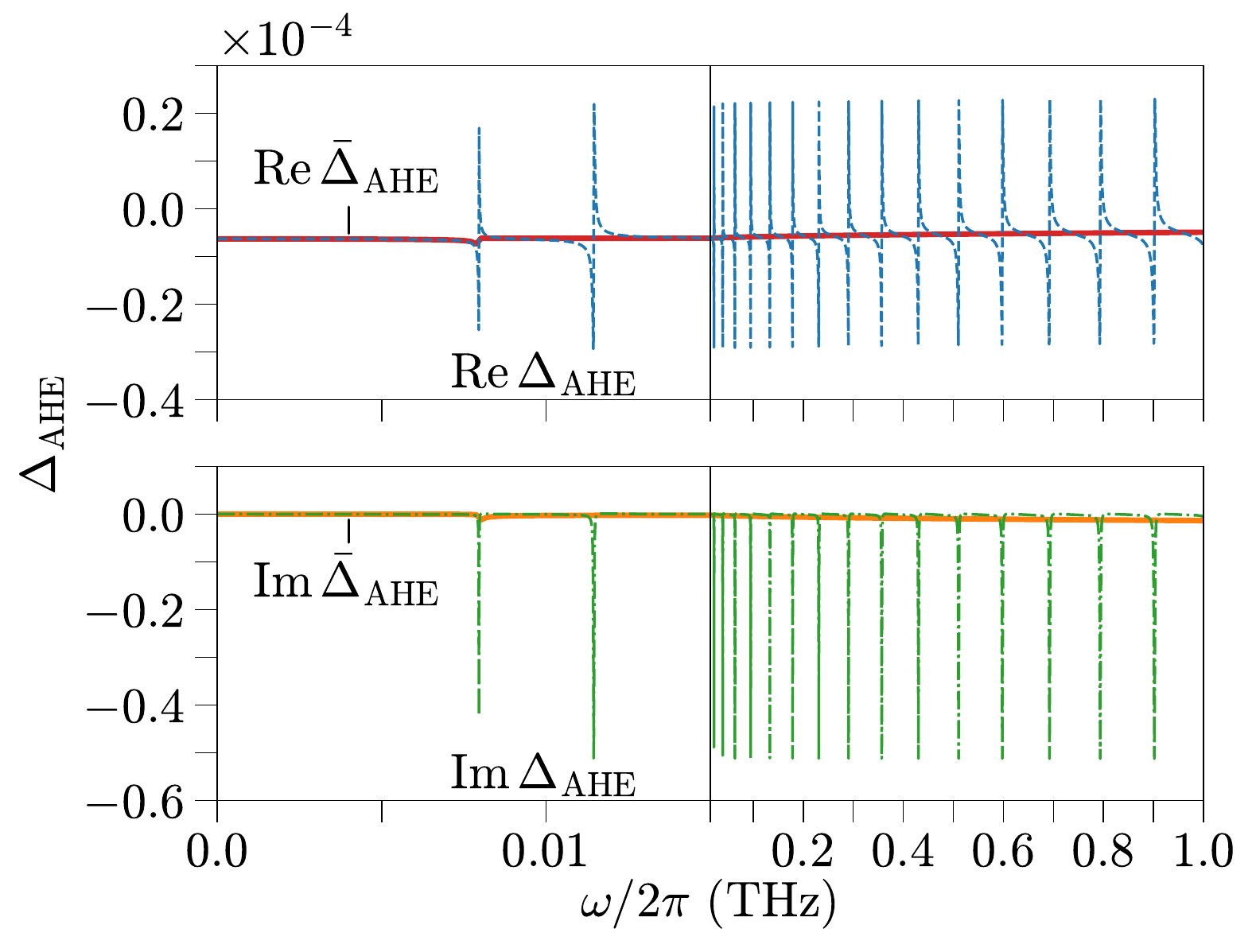}
\caption{Real and imaginary parts of the dimensionless characteristics $\Delta_{\rm SMR}$ of the spin-Hall magnetoresistance (top) and $\Delta_{\rm AHE}$ of the anomalous Hall effect (bottom) for a YIG$|$Pt bilayer. The thick solid lines use the asymptotic value $Z_{{\rm F}\perp}^{\infty}(\omega)$ of Eq. \eqref{eq:Zinfinity}; thin dashed lines use the full expression for $Z_{{\rm F}\perp}(\omega)$, which includes the effect of spin-wave resonances in the YIG layer. Material and device parameters are taken from Tab.\ \ref{tab:estimates_parameters}. \label{fig:YIG|Pt_SMR_ratios_re_im}}
\end{figure}

\begin{figure}
\centering
\includegraphics[width=0.5\textwidth]{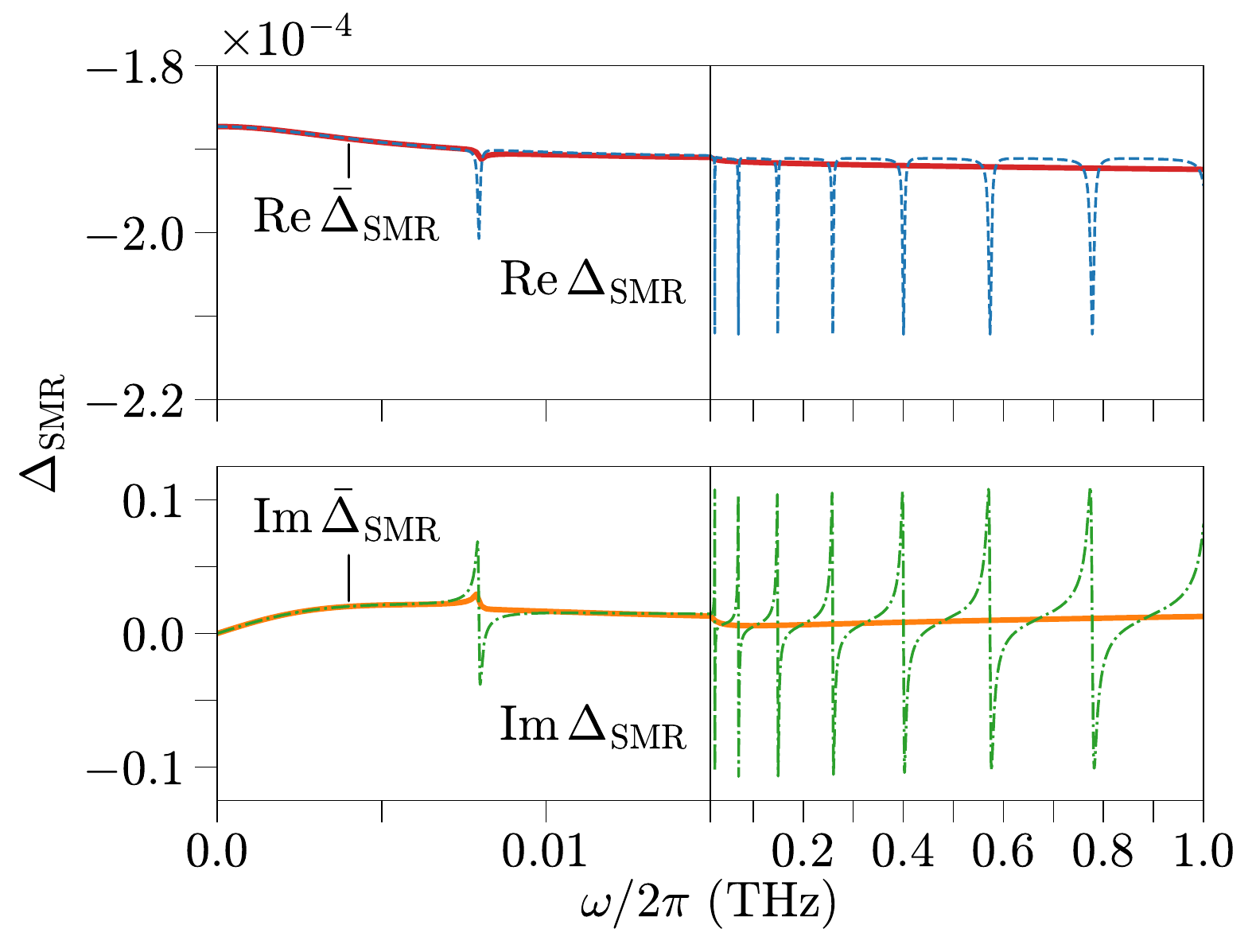}

\includegraphics[width=0.5\textwidth]{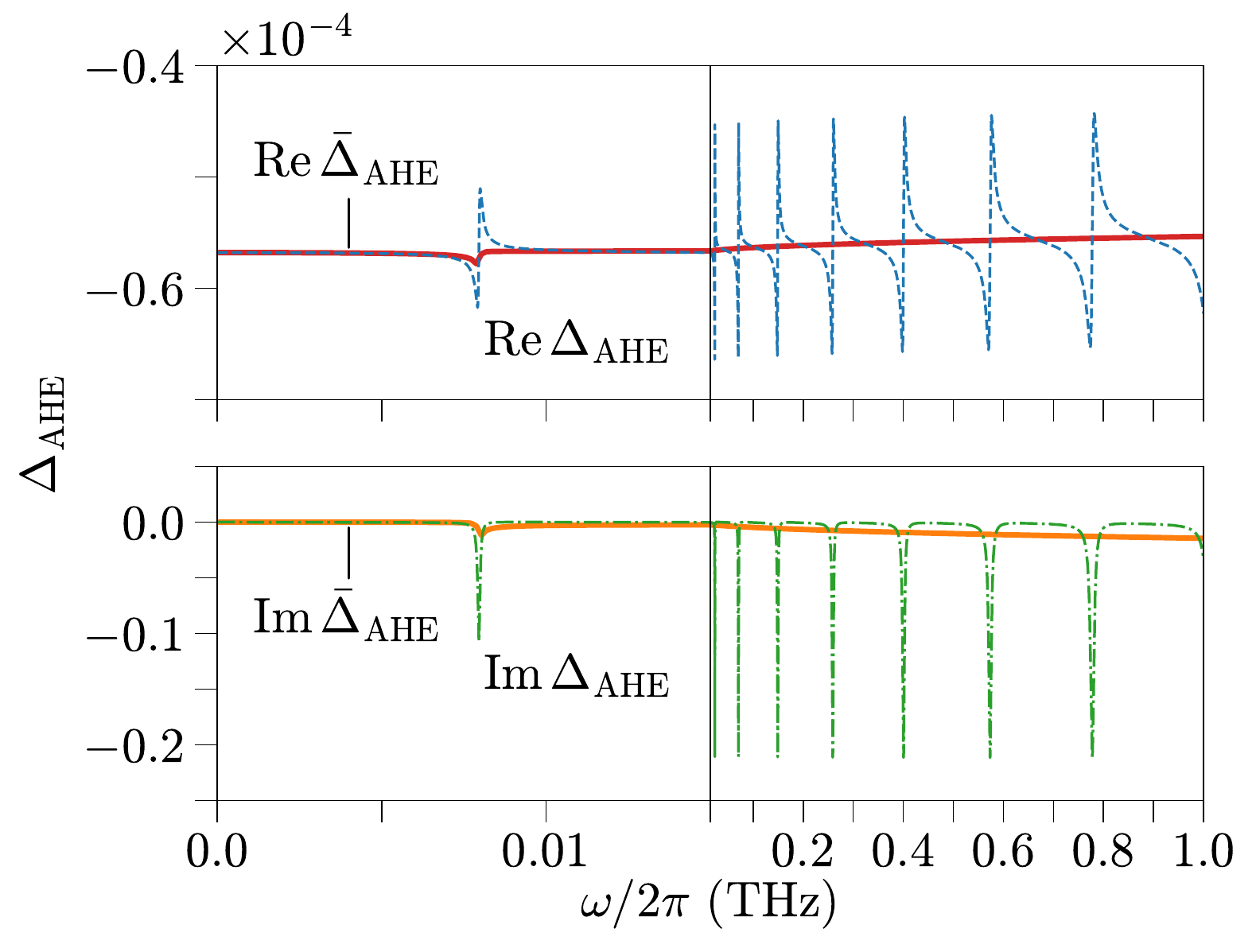}
\caption{Same as Fig. \ref{fig:YIG|Pt_SMR_ratios_re_im}, but for an Fe$|$Au bilayer. Parameter values are taken from Tab.\ \ref{tab:estimates_parameters}.\label{fig:Fe|Au_SMR_ratios_re_im}}
\end{figure}

Below we discuss the full frequency-resolved characteristics $\Delta_{\rm SMR}$ and $\Delta_{\rm AHE}$, which contain a contribution from the sharp magnon resonances in F, as well as from the asymptotic characteristics $\bar \Delta_{\rm SMR}$ and $\bar \Delta_{\rm AHE}$, which are obtained by replacing the transverse spin impedance $Z_{{\rm F}\perp}(\omega)$ with $Z_{{\rm F}\perp}^{\infty}(\omega)$ of Eq.\ \eqref{eq:Zinfinity}. For definiteness, we take experiment-motivated sample thicknesses $d_{\rm N}$ and $d_{\rm F}$ as listed in Tab. \ref{tab:estimates_parameters}. Numerical estimates for $\Delta_{\rm SMR}$ and $\Delta_{\rm AHE}$ are shown in Figs.\ \ref{fig:YIG|Pt_SMR_ratios_re_im} and \ref{fig:Fe|Au_SMR_ratios_re_im} for YIG$|$Pt and Fe$|$Au bilayers, respectively. No separate results are shown for $\Delta_{\rm PHE}$, since $\Delta_{\rm PHE} = (1/2) \Delta_{\rm SMR}$, see Eq.\ (\ref{eq:DeltaPHE}).

In both figures the frequency dependence of $\Delta_{\rm SMR}$ at low frequencies ($\omega/2 \pi \lesssim 1\, {\rm GHz}$) mainly results from the frequency dependence of $Z_{{\rm F}\parallel}^{\rm m} (\omega)$, while above $\omega_0$ it stems from the resonance structure of $Z_{{\rm F}\perp}(\omega)$. Comparing $\mbox{Re}\, \Delta_{\rm SMR}(\omega)$ at $\omega = 0$ and $\omega/2 \pi = 1\, {\rm THz}$, there is a decrease of the size of the SMR by 68\% for YIG$|$Pt and an increase by ca.\ 3\% for Fe$|$Au, which mainly has its origin in the frequency range below $1\, {\rm GHz}$. The net frequency dependence of $\Delta_{\rm SMR}$ for Fe$|$Au is weaker than for YIG$|$Pt, because for Fe$|$Au the frequency-independent electronic spin impedances $Z_{\rm FN \parallel}^{\rm e} + Z_{\rm F \parallel}^{\rm e}$ shunt the frequency-dependent magnon spin impedances $Z_{\rm FN \parallel}^{\rm m} + Z_{\rm F \parallel}^{\rm m}(\omega)$ (cf. Fig. \ref{fig:equiv_circuit}). For Fe$|$Au, $\Delta_{\rm SMR}$ is negative, as the longitudinal contributions dominate over the transverse one.
In Fig.\ \ref{fig:YIG|Pt_SMR_ratios_re_im}, the imaginary part ${\rm Im}\,  \Delta_{\rm SMR}$ exhibits a peak for frequencies $\omega/2 \pi$ in the GHz range, where the (almost purely imaginary) longitudinal spin impedance $Z_{{\rm F}\parallel}^{\rm m}$ matches the corresponding interface impedance $Z_{{\rm FN}\parallel}^{\rm m}$.

The discussion of $Z_{{\rm F}\perp}^{\infty}(\omega)$ in Sec.\ \ref{subsec:llg} implies that the asymptotic characteristic $\bar \Delta_{\rm SMR}$ describes the SMR for the case that the frequency resolution is larger than the spacing between spin-wave resonances in F. Although the spin impedance $Z_{{\rm F}\perp}^{\infty}(\omega)$ depends strongly on frequency, see Sec.\ \ref{subsec:llg}, for both material combinations the transverse contribution to $\bar \Delta_{\rm SMR}$ depends only very weakly on frequency, as the spin impedance $Z_{{\rm F}\perp}^{\infty}(\omega)$ appears in series with the much larger interface spin impedance $Z_{{\rm FN}\perp}$, which is frequency independent.
The same applies to the PHE and AHE corrections to the conductivity, characterized by $\bar \Delta_{\rm PHE}$ and $\bar \Delta_{\rm AHE}$, respectively.

If the effect of a finite thickness $d_{\rm F}$ is taken into account, the SMR response acquires sharp resonances, reflecting the resonance structure of $Z_{{\rm F}\perp}(\omega)$. The real part of $\Delta_{\rm SMR}$ shows narrow symmetric features for frequencies around the standing spin-wave modes in F. For a quantitative discussion we notice that the transverse contribution to the conductivity correction $\delta \sigma^{xx}(\omega)$ is a sum of contributions involving impedances at frequencies $\omega$ and $-\omega$. Of these, it is only the positive-frequency contribution that is affected by the resonances in $Z_{{\rm F}\perp}(\omega)$. Neglecting the small imaginary part of $Z_{{\rm FN}\perp}$ and taking into account that $|Z_{{\rm F}\perp}^{\infty}(\omega)| \ll |Z_{{\rm FN}\perp}|$ for both material combinations we consider, one finds that at the resonance center the transverse contribution to $\Delta_{\rm SMR}$ is reduced by a factor $\approx [1 - d_{\rm F}^{\rm c}/2(d_{\rm F} + d_{\rm F}^{\rm c})]$, with $d_{\rm F}^{\rm c} = 2 \gamma \hbar/\alpha M_{\rm s} (Z_{\rm N} + Z_{{\rm FN}\perp})$. The full width at half maximum of the resonant features in $\mbox{Re}\, \Delta_{\rm SMR}$ is $\delta \omega \approx 2 \alpha \omega (1 + d_{\rm F}^{\rm c}/d_{\rm F})$. The crossover scale $d_{\rm F}^{\rm c}$ separates the low-$d_{\rm F}$ regime,\cite{Tserkovnyak_2002_a,Tserkovnyak_2002_b} in which the lifetime of spin waves is limited by decay into N, and the large-$d_{\rm F}$ regime, in which intrinsic Gilbert damping determines the magnon lifetime. Taking material parameters from Tab.\ \ref{tab:estimates_parameters}, we find that $d_{\rm F}^{\rm c} \approx 0.2\, \mu{\rm m}$ for YIG$|$Pt and $d^{\rm c}_{\rm F} \approx 0.6\, {\rm nm}$ for Fe$|$Au. The imaginary part of $\Delta_{\rm SMR}$ has abrupt jumps at the spin-wave frequencies and is a smooth function of frequency otherwise. The same discussion applies to $\Delta_{\rm PHE}$ and to $\Delta_{\rm AHE}$, but with the roles of real and imaginary parts reversed for the latter. 

Lotze {\it et al.}\cite{Lotze_2014} performed a measurement of the spin-Hall magnetoresistance of a YIG$|$Pt bilayer for frequencies up to $\omega/2\pi = 3\,{\rm GHz}$. The magnetic field in this experiment was $0.6\,{\rm T}$, resulting in a ferromagnetic-resonance frequency $\omega_0/2 \pi \approx 17\,{\rm GHz}$, which is an order of magnitude larger than what we used for our numerical evaluation. The increased value of $\omega_0$ shifts the lowest magnon resonances to higher frequencies and it lowers the ``magnon capacitance'' $C_{\rm m}$, see Eq. \eqref{eq:Cm}. As a result, the low-frequency regime, in which $Z_{{\rm F}\parallel}^{\rm m}(\omega)$ dominates the longitudinal spin current, is extended to higher frequencies. Setting $\omega_0/2 \pi = 17\, {\rm GHz}$ and taking the other material parameters from Tab.\ \ref{tab:estimates_parameters}, we find that ${\rm Re}\, \Delta_{\rm SMR}$ decreases by approximately $10 \%$
between $\omega = 0$ and $\omega/2 \pi = 3\,{\rm GHz}$, consistent with the observation of Ref.\ \onlinecite{Lotze_2014}, who found that the magnitude of the spin-Hall magnetoresistance does not change within the experimental uncertainty of $\sim 10 \%$.

\section{Conclusion and Outlook}
\label{sec:conclusions}

The main difference between the zero-frequency and finite-frequency versions of the spin-Hall magnetoresistance (SMR) results from the onset of spin transport by incoherent magnons and the resonant coherent excitation of magnetization modes in F for frequencies in the GHz range and above.
Extending the work of Chiba {\em et al.},\cite{Chiba_2014} who developed a theory of the finite-frequency SMR that accounted for the uniform precession mode at the ferromagnetic-resonance frequency,
we here present a theory that includes acoustic spin waves of arbitrary wavelength, accounting for their role in incoherent longitudinal as well as coherent transverse spin transport, thus allowing for an extension to the GHz and THz frequency range.
Our theory is a general linear response theory, in which the relevant response functions of the normal metal, the ferromagnet, and the interface are lumped together into ``spin impedances'', see Fig. \ref{fig:equiv_circuit}. These spin impedances relate spin current and spin accumulation in the same way as standard impedances relate charge current and voltage. This allows for an efficient description of F$|$N bilayers and F$|$N$|$F trilayers, in which the thickness $d_{\rm N}$ exceeds the spin-relaxation length $\lambda_{\rm N}$ --- so that the SMR is the sum of contributions from the two interfaces of the normal-metal layer ---, as well as of bilayers of a smaller thickness.

The SMR corrections $\delta \sigma^{xx}$ and $\delta \sigma^{xy}$ to the diagonal and off-diagonal conductivity --- see Eqs. \eqref{eq:jcx2}, \eqref{eq:jcy2} and Figs. \ref{fig:YIG|Pt_SMR_ratios_re_im}, \ref{fig:Fe|Au_SMR_ratios_re_im} --- are the difference of contributions associated with the flow of spin angular momentum collinear (``longitudinal'') and perpendicular (``transverse'') to the magnetization direction. The longitudinal contribution is carried by conduction electrons and thermal magnons inside the ferromagnet. The magnonic longitudinal contribution to the SMR, which is the sole longitudinal contribution if F is insulating, has a systematic frequency dependence, with a characteristic frequency scale in the GHz range for magnetic layer thickness $d_{\rm F} \sim 10\, {\rm nm}$. The origin of this frequency dependence is the build-up of a magnon chemical potential in F, which effectively blocks a longitudinal magnonic spin current in the DC limit, but not for high frequencies. The transverse contribution, which is carried by coherent magnons, features sharp resonances at the spin-wave frequencies of the F$|$N bilayer, but otherwise has only a small systematic frequency dependence. With sufficient frequency resolution the THz-SMR may be an interesting tool for all-electric spectroscopy of magnon modes. Furthermore, our analysis has shown that the excitation and propagation of coherent magnons is as efficient at THz frequencies as in the GHz range.

The combination of the longitudinal and transverse contributions leads to a significant decrease of the SMR for YIG$|$Pt by $68 \%$ between $\omega = 0$ to $\omega/2\pi = 1\,$THz. The SMR decreases so strongly, because for bilayers involving YIG as a ferromagnetic insulator, the transverse and longitudinal contributions to the SMR are of comparable magnitude at high frequencies at room temperature --- with the transverse contribution dominating ---, whereas the longitudinal contribution is suppressed at low frequencies. The thickness $d_{\rm F}$ of the ferromagnetic layer, the applied magnetic field, and the choice of the normal metal affect the characteristic crossover frequency, but not the over-all change of the SMR between the low and high frequency limits. Since the ratio of longitudinal to transverse spin impedances of the F-N interface systematically decreases with temperature,\cite{Bender_2015,Cornelissen_2016,Goennenwein_2015,Wu_2016} we expect that the longitudinal and transverse contributions for bilayers involving YIG as a ferromagnetic insulator eventually cancel each other and, hence, that the SMR vanishes, upon raising the temperature above room temperature. Further, whereas for the thicknesses typically used in experiments the characteristic frequency for the onset of the longitudinal contribution to the SMR is comparable to the ferromagnetic-resonance frequency $\omega_0$, the two frequency scales can be pushed apart by considering smaller or larger layer thicknesses. In particular, for large $d_{\rm F}$ we predict that the frequency-dependent suppression of the SMR can set in significantly below $1\, {\rm GHz}$, so that it is measurable by conventional high-frequency electronic techniques.

An upper limit for the applicability of our theory is the maximum frequency $\omega_{\rm max}$ of acoustic magnons. For the magnets we consider here, one has \mbox{$\omega_{\rm max}/2 \pi \sim 5\, {\rm THz}$}.\cite{Cherepanov_1993,Barker_2016} At this frequency, both the description of the F-N interface as a mere ``boundary condition'' and the description of magnetization dynamics by the Landau-Lifshitz-Gilbert equation (\ref{eq:llg}) cease to be valid. Approximately the same frequency restricts the use of the quasi-adiabatic approximation for longitudinal spin transport by thermal magnons at room temperature. Both frequency limits are below frequency scales at which other assumptions of our theory cease to be valid, such as neglecting frequency-dependent corrections to the Drude scattering rates in N and F (valid for $\omega/2 \pi \lesssim 10\, {\rm THz}$).

Apart from the appearance of sharp resonances at spin wave frequencies, the transverse contribution to the SMR has no appreciable systematic frequency dependence for the material combinations YIG$|$Pt and Fe$|$Au for frequencies well into the THz regime. Within our formalism, the reason is that the transverse contribution to the SMR is dominated by the frequency-independent spin impedance $Z_{{\rm FN}\perp}$ of the interface, which obscures the strong frequency dependence of $Z_{{\rm F}\perp}(\omega)$ in the THz regime. We have also considered other materials that are used in spintronics experiments, such as Cu, Co, CoFe(B), Py (permalloy), GdFe(Co), GdIG (gadolinium iron garnet), Fe$_3$O$_4$ (magnetite), NiFe$_2$O$_4$, EuS \cite{Gomez-Perez_2020}, and EuO \cite{Rosenberger_2021} in combination with different non-magnetic heavy metals as N layers and arrived at the same qualitative conclusion: In all cases, the transverse impedance $Z_{{\rm F}\perp}^{\infty}(\omega)$, which describes the frequency-averaged part of magnon-mediated spin transport in F, is much smaller than $Z_{{\rm FN}\perp}$, ruling out a substantiative effect of the magnon spin impedance $Z_{{\rm F}\perp}^{\infty}$ on the SMR.

This leads to the question, whether there are other, less-explored material combinations, for which the transverse contribution to the SMR has a stronger systematic frequency dependence. To address this question, we find it instructive to consider $Z_{{\rm F}\perp}(\omega)$ at the highest frequency $\omega_{\rm max}$ for which our long-wavelength theory is valid, which is at the boundary of the (magnetic) Brillouin zone. Using $M_{\rm s}/\hbar \gamma = S/a_{S}^x a_{S}^y a_{S}^z$, where $S$ and $a_{S}^{x,y,z}$ are the spin and linear dimensions of the magnetic unit cell, respectively, and $\omega_{\rm max} \approx \Dex (\pi/a_{z})^2$, we estimate that
\begin{equation}
  |Z_{{\rm F}\perp}^{\infty}(\omega)| \le |Z_{{\rm F}\perp}(\omega_{\rm max})| \sim \frac{\pi a_{S}^x a_{S}^y}{S}.
  \label{eq:ZFestimate}
\end{equation}
A lower limit for the interface impedance $Z_{{\rm FN}\perp}$ is obtained using the Sharvin resistance of the F-N interface,\cite{Brataas_2000,Zwierzycki_2005}
\begin{equation}
  Z_{{\rm FN}\perp} \gtrsim 4 \pi \lambda_{\rm e}^2,
  \label{eq:ZFNestimate}
\end{equation}
where $\lambda_{\rm e}$ is the Fermi wavelength in N. The fact that for most material combinations $\lambda_{\rm e}$ and $a_{S}^{x,y}$ are comparable, whereas $S$ is of order unity or larger, explains why material combinations with $Z_{{\rm F}\perp}(\omega)$ larger than $Z_{{\rm FN}\perp}$ are hard to find.

As a guiding principle for the search for material combinations in which $Z_{{\rm F}\perp}(\omega)$ is large and $Z_{{\rm FN}\perp}$ is small, Eqs.\ (\ref{eq:ZFestimate}) and (\ref{eq:ZFNestimate}) suggest to consider materials with a large magnetic unit cell, small $S$, and a spin-mixing conductance that approaches the Sharvin limit as closely as possible. The spin stiffness $\Dex$ does not directly enter into the comparison of $Z_{{\rm F}\perp}(\omega_{\rm max})$ and $Z_{{\rm FN}\perp}$, but a small $\Dex$ lowers the frequency scale $\omega_{\rm max}$, making it easier to reach this frequency scale experimentally. A promising class of materials in this regard are half-metallic (fully) compensated ferrimagnetic Heusler compounds, such as Mn$_3$Al or Mn$_{1.5}$FeV$_{0.5}$Al. The former compound has a band gap $\sim 0.5\, {\rm eV}$ for minority carriers\cite{Han_2019} and a magnetization $M_{\rm s} \approx 2 \times 10^{4}\, {\rm Am}^{-1}$,\cite{Jamer_2017} which is almost an order of magnitude below the corresponding value for YIG. The small value of $M_{\rm s}$ should result in a relatively large value of $Z_{{\rm F}\perp}$ --- consistent with the expectation that acoustic magnons do not efficiently transport spin angular momentum in an almost compensated ferrimagnet ---, whereas the half-metallic character ensures a large spin-mixing conductance. Indeed, the related half-metallic ferromagnetic Heusler compounds such as Co$_2$MnSi\cite{Carva_2007,Carva_2007_2,Turek_2007,Chudo_2011,Sasaki_2020}, Co$_2$FeAl\cite{Kumar_2019}, and Co$_2$Fe$_{0.4}$Mn$_{0.6}$Si\cite{Singh_2019} are reported to have spin-mixing conductances around or above the spin-mixing conductance of YIG$|$Pt. For a more complete answer to this question, however, a detailed calculation of the spin-mixing conductances for the almost compensated ferrimagnetic compounds is necessary, as well as a description of the F-N interface that accounts for the effect of the finite penetration of minority electrons into F on the coupling to short-wavelength acoustic magnons.

\section*{Acknowledgements}

We thank Y. Acremann, G. Bierhance, S. T. B. G\"onnenwein, L.\ N\'advornik, M. A. Popp, I. Radu, U. Ritzmann, S. M. Rouzegar, and R. Schlitz for stimulating discussions. This work was funded by the DFG via the collaborative research center SFB-TRR 227 ``Ultrafast Spin Dynamics" (projects A05, B02, and B03).

\begin{appendix}

\section{Longitudinal spin current for ballistic magnon dynamics}
\label{subsec:ferromagnet_long_mag_ballistic}

In this appendix we describe the longitudinal magnonic spin current through the F-N interface if the magnon dynamics in F is ballistic, with specular reflection at the interface of the F layer to the vacuum at $z = -d_{\rm F}$. 
Magnons at the F-N interface are described by their distribution function $f(\Omega,k_x,k_y)$, explicitly accounting for the wavevector components $k_x$ and $k_y$ parallel to the F-N interface. We distinguish the distribution functions $f_{\rm in}(\Omega,k_x,k_y)$ of magnons incident on the F-N interface and $f_{\rm out}(\Omega,k_x,k_y)$ of magnons moving away from the interface. With ballistic magnon transport in F and specular reflection of magnons at the ferromagnet-vacuum interface at $z=-d_{\rm F}$, we obtain the relation
\begin{equation}
  f_{\rm in}(\Omega,k_x,k_y;t) = f_{\rm out}(\Omega,K_x,K_y;t-2 d_{\rm F}/v_z),
  \label{eq:fcond}
\end{equation}
where $v_z(\Omega,k_x,k_y) = 2 \Dex k_z (\Omega,k_x,k_y)$ is the magnon velocity perpendicular to the interface.

Magnons incident on the F-N interface are reflected into F with probability $R_{\rm m}(\Omega,k_x,k_y)$. The difference $T_{\rm m} = 1 - R_{\rm m}$ is the probability that a magnon is annihilated at the interface and transfers its angular momentum to the conduction electrons in N. It is\cite{Schmidt_2021,Reiss_2021}
\begin{equation}
  T_{\rm m}(\Omega,k_x,k_y) = \frac{M_{\rm s} \Dex k_z \Omega/\pi \hbar \gamma}{|M_{\rm s} \Dex k_z/\hbar \gamma + \Omega g_{\uparrow\downarrow}/4 \pi|^2}\, \mbox{Re}\, g_{\uparrow\downarrow},
\end{equation}
where
\begin{equation}
  k_z = \sqrt{\frac{\Omega - \omega_0}{\Dex} - k_x^2 - k_y^2}.
\end{equation}
The magnon distribution at the interface then satisfies the additional boundary condition\cite{Schmidt_2021}
\begin{align}
  f_{\rm out}(\Omega,k_x,k_y;t) =&\, R_{\rm m}(\Omega,k_x,k_y) f_{\rm in}(\Omega,k_x,k_y;t)
\label{eq:boundary}   \\ \nonumber + & T_{\rm m}(\Omega,k_x,k_y) f^0(\Omega - \mu_{{\rm s}\parallel}(z \downarrow 0,t)/\hbar),
\end{align}
where $f^0$ is the Planck distribution. The second term in Eq.\ (\ref{eq:boundary}) ensures that $f_{\rm out} = f_{\rm in}$ if the temperatures in F and N are the same and the magnon chemical potential in F equals the spin accumulation in N.\cite{Cornelissen_2016} 

\begin{figure}
\centering
\includegraphics[width=0.5\textwidth]{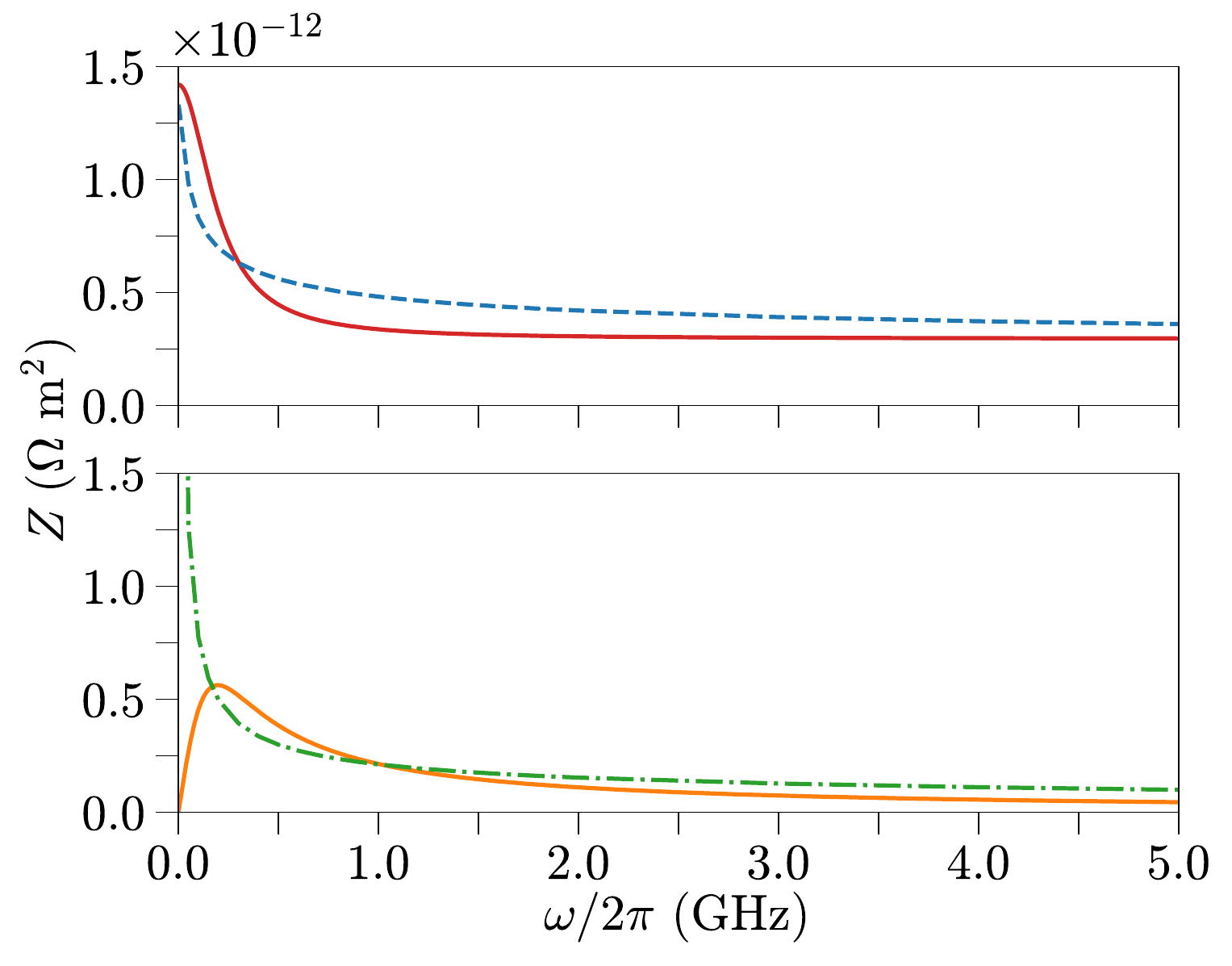}
\caption{\label{fig:ballistic} Real part (top) and imaginary part (bottom) of the combined impedance $Z_{\parallel}^{\rm m}(\omega)$ from Eq.\ (\ref{eq:zexpressionfinite}) for a ballistic F layer (dashed and dot-dashed curve) and of the impedance sum $Z_{{\rm FN}\parallel}^{\rm m} + Z_{{\rm F}\parallel}^{\rm m}(\omega)$ (solid curves) of Secs. \ref{sec:trans+long_SMR} and \ref{subsec:ferromagnet_long_mag}, \ref{subsec:ferromagnet_long_mag_diffusive}. Material and device parameters are taken from Tab.\ \ref{tab:estimates_parameters}.}
\end{figure}

In linear response, the distribution functions $f_{\rm in}$ and $f_{\rm out}$ may be expanded as
\begin{align}
  f_{\rm in,out}(\Omega,k_x,k_y;t) =&\, f^0(\Omega) \\ \nonumber &\, \mbox{} + \left( - \frac{df^0(\Omega)}{d\Omega} \right)
  \phi_{\rm in,out}(\Omega,k_x,k_y;t).
\end{align}
Setting $\mu_{{\rm s}\parallel}(t) = \mu_{{\rm s}\parallel}(\omega) e^{-i \omega t} + \mu_{{\rm s}\parallel}(- \omega) e^{+ i \omega t}$ and solving for the linear-response corrections $\phi_{\rm in,out}(\Omega,k_x,k_y;t)$ to the distribution functions, we find that the spin current through the F-N interface is
\begin{equation}
  j_{{\rm sm}\parallel}(0, \omega) = -\frac{1}{Z^{\rm m}_{\parallel}(\omega)} \mu_{{\rm s}\parallel}(z \downarrow 0, \omega),
\end{equation}
with
\begin{align}
  \frac{1}{Z^{\rm m}_{\parallel}(\omega)} =&\, \frac{1}{(2 \pi)^3}
  \int dk_x dk_y
  \int d\Omega\,
  \left( - \frac{d f^0}{d\Omega} \right) T_{\rm m}(\Omega, k_x,k_y) 
  \label{eq:zexpressionfinite} \\ 
  &\, \mbox{} \times
  \left[1 - 
  \frac{T_{\rm m}(\Omega,k_x,k_y) e^{2 i \omega d_{\rm F}/v_z}}
       {1 - R_{\rm m}(\Omega,k_x,k_y) e^{2 i \omega d_{\rm F}/v_z}} \right].
\nonumber
\end{align}
The first term between the square brackets is identical to the (inverse) longitudinal interfacial spin impedance $Z_{{\rm FN}\parallel}^{\rm}$. The correction term accounts for the magnon accumulation in F. In the limit of zero frequency, the correction term imposes that $j_{{\rm sm}\parallel} \to 0$, {\em i.e.}, $Z_{\parallel}^{\rm m} \to \infty$. For frequencies large enough such that $\omega d_{\rm F} \gg v_z$ for thermal magnons, the integrand in the correction term is a fast-oscillating function of frequency, so that $Z_{\parallel}^{\rm m}(\omega) \to Z_{{\rm FN}\parallel}^{\rm m}$. Fig. \ref{fig:ballistic} compares the ballistic impedance $Z_{\parallel}^{\rm m}(\omega)$ of Eq.\ (\ref{eq:zexpressionfinite}) with the corresponding sum impedance \mbox{$Z_{{\rm FN}\parallel}^{\rm m} + Z_{{\rm F}\parallel}^{\rm m}(\omega)$} of a YIG$|$Pt interface calculated using the effective magnetoelectronic circuit theory of the main text. The plot illustrates this limiting behavior of $Z_{\parallel}^{\rm m}(\omega)$.

\section{Response to Oersted field}
\label{sec:magnonic_spin_F}

The (Oersted) magnetic field of the alternating charge current $\vj^{\rm c}(t)$ drives a magnetization precession which, via spin pumping and the ISHE, results in an additional correction to the charge current. Since the SMR correction $\delta \vj^{\rm c}$ to the charge current density in N is small, we may neglect it when calculating the magnetic field $\vB(t)$ in F and set\footnote{In addition to the Oersted field there is also the magnetic field associated with the driving THz radiation. For the material and device parameters used here, see Tab.\ \ref{tab:estimates_parameters}, one has $B_y/E = \mu_0 \sigma_{\rm N} d_{\rm N}/2 \approx 2 \times 10^{-8}\, \rm{s\, m}^{-1}$ for YIG$|$Pt, and $\approx 2 \times 10^{-6} \rm{s\, m}^{-1}$ for Fe$|$Au. In contrast, for the direct magnetic field of the driving radiation one has $B_y/E = 1/c \approx 3 \times 10^{-9}\, \rm{s\, m}^{-1}$. Inclusion of the direct magnetic field into the considerations of App.\ \ref{sec:magnonic_spin_F} is straightforward, if desired.}
\begin{equation}
  \vB(t) = B_y(t) \ve_y,\ \ B_y(t) = \mu_0 \frac{E(t) \sigma_{\rm N} d_{\rm N}}{2}.
\end{equation}
The magnetization response is found by inclusion of the magnetic field into the Landau-Lifshitz-Gilbert equation (\ref{eq:llg}), 
\begin{align}
  \label{eq:llg2}
  \dot \vm =&\, \omega_0 \, \ve_{\parallel} \times \vm + \alpha \, \vm \times \dot \vm - \Dex \, \vm \times \frac{\partial }{\partial z^2} \vm
  \nonumber \\ &\, \mbox{}
  + \gamma \, \vm \times \vB.
\end{align}
Linearizing Eq.\ (\ref{eq:llg2}) and performing a Fourier transformation to frequency as in Sec.\ \ref{subsec:llg}, we find that only the transverse component
\begin{align}
  B_{\perp}(\omega) =&\, \ve_{\perp}^* \cdot \vB(\omega)
\end{align}
of the Oersted field affects the magnetization dynamics,
\begin{align}
  -\Dex
    \frac{\partial^2 }{\partial z^2} m_{\perp}(z,\omega)
  =&\,
  (\omega + i \alpha \omega - \omega_0) m_{\perp}(z,\omega)
  \nonumber \\ &\, \mbox{}
  - \gamma B_{\perp}(\omega).
\end{align}
With the boundary condition that the spin currents must vanish at the boundary of the F layer at $z=-d_{\rm F}$, the solution of this equation is
\begin{align}
  m_{\perp}(z,\omega) =&\, m_{\perp}(0,\omega)
  \frac{\cos[K(\omega) (z + d_{\rm F})]}{\cos[K(\omega) d_{\rm F}]}
    \nonumber \\ &\,
  + \chi_{\perp}(\omega) B_{\perp}(\omega),
\end{align}
where
\begin{equation}
  \chi_{\perp}(\omega) = \frac{\gamma}{\omega + i \alpha \omega - \omega_0}.
\end{equation}
The uniform magnetization precession driven by the Oersted field does not carry a spin current inside F, but it does lead to an additional contribution to the spin current through the F-N interface via spin pumping. Hence,
Eqs. \eqref{eq:ZN1} and \eqref{eq:ZF1} remain unchanged, whereas the transverse part of Eqs. \eqref{eq:ZFN1} has to be modified,
\begin{align}
  \mu_{{\rm s}\perp}(z \downarrow 0,\omega) +
  \hbar \omega m_{\perp}(0,\omega) =&\,
  - Z_{{\rm FN}\perp}(\omega) j_{{\rm s}\perp}^z(0,\omega) \nonumber\\
  &\, - \chi_{\perp}(\omega) \hbar \omega B_{\perp}(\omega).
\end{align}
Repeating the calculations of Sec.\ \ref{sec:trans+long_SMR}, we then find that the corrections $\delta \sigma^{xx}$ and $\delta \sigma^{xy}$ to the in-plane conductivity are still given by Eq.\ (\ref{eq:jcx2}) and (\ref{eq:jcy2}), but with the replacement
\begin{align}
\label{eq:Oersted_factor}
  \frac{1}{Z_{\perp}(\omega)} \to
  \frac{1}{Z_{\perp}(\omega)} \left[1 + \frac{e \omega \mu_0 \chi_{\perp}(\omega) d_{\rm N}}{Z_{\rm N} \theta_{\rm SH}} \right].
\end{align}

\begin{figure}
\centering
\includegraphics[width=0.5\textwidth]{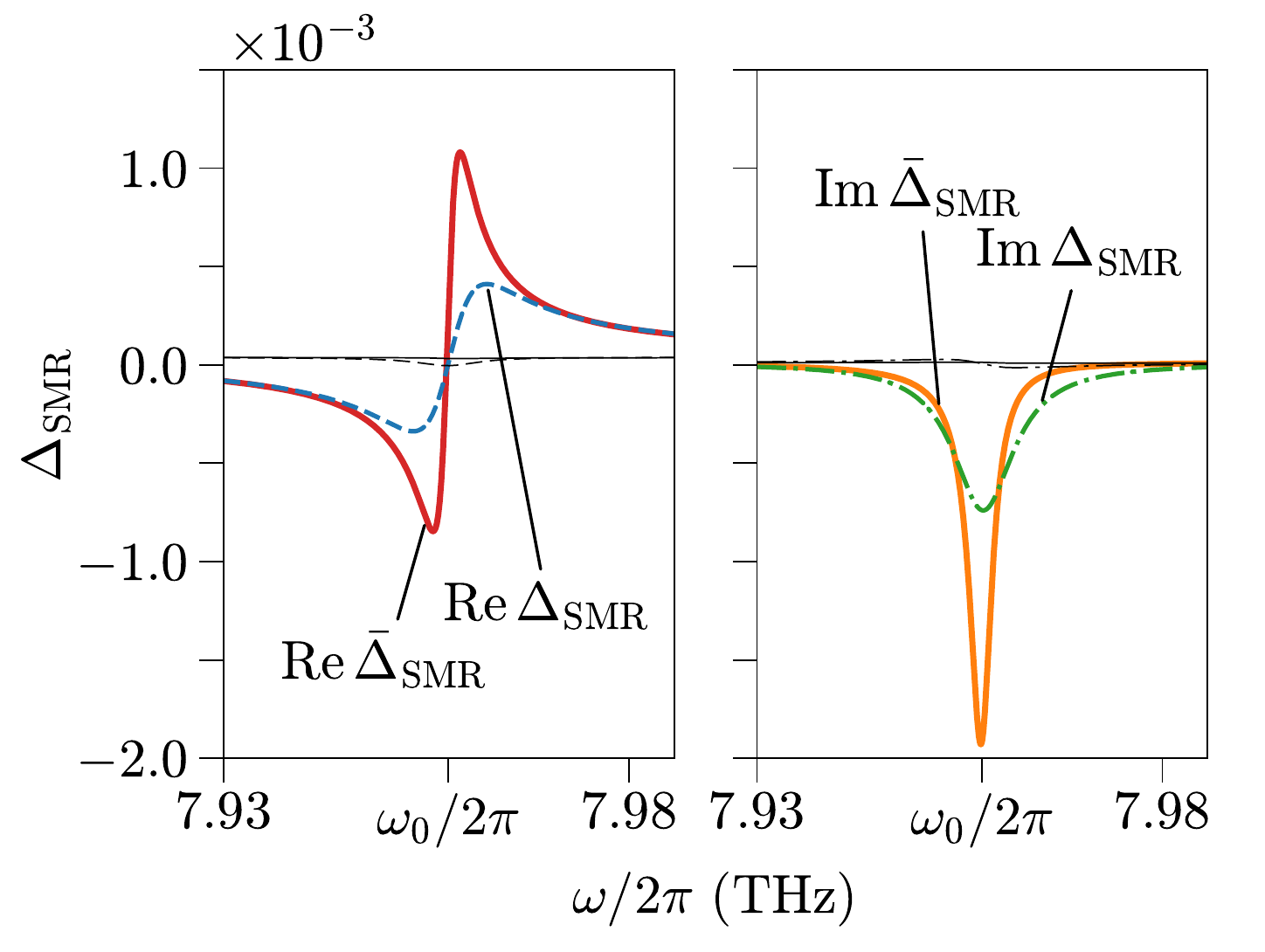}
\caption{Comparison of the characteristecs $\Delta_{\rm SMR}$ (dashed and dot-dashed curve) and $\bar \Delta_{\rm SMR}$ (solid curves) with and without inclusion of the Oersted field. Thick (colored) curves are for the case with Oersted field, thin (black) curves are for the case without it. The left and right panels show real and imaginary parts, respectively. Parameter values are taken from Tab.\ \ref{tab:estimates_parameters}.
\label{fig:YIG|Pt_smr_w_wo_oersted}}
\end{figure}

For the material and device parameters used in Secs.\ \ref{subsec:impedances_numerical_estimates} and \ref{sec:experiments}, see Tab.\ \ref{tab:estimates_parameters}, inclusion of the Oersted field only affects the conductivity correction for frequencies in the immediate vicinity of the ferromagnetic-resonance frequency $\omega_0$, where the transverse susceptibility $\chi_{\perp}(\omega)$ is maximal. This is illustrated in Fig.\ \ref{fig:YIG|Pt_smr_w_wo_oersted} for the case of a YIG$|$Pt bilayer. Inclusion of the Oersted field has a much larger effect on the SMR in the limit of large $d_{\rm F}$ (solid curves in Fig.\ \ref{fig:YIG|Pt_smr_w_wo_oersted}) than in the limit of small $d_{\rm F}$ (dashed and dash-dotted curves). Also, upon inclusion of the Oersted field, the form of the lineshape of the resonance near $\omega_0$ is switched between imaginary and real parts of $\Delta_{\rm SMR}$ --- something that can be understood noting that the correction factor in Eq.\ (\ref{eq:Oersted_factor}) is almost purely imaginary for $\omega = \omega_0$.
Since the Oersted field only drives a uniform precession mode of the magnetization, it does not couple to the magnon modes at higher frequencies.

\section{SMR for F$|$N bilayers and F$|$N$|$F trilayers}
\label{sec:spin_valves}

The results of Sec.\ \ref{sec:trans+long_SMR} describe the corrections to the charge current densities $j_{\rm c}^x$ and $j_{\rm c}^y$ associated with a single interface for the case that the normal layer has a thickness $d_{\rm N}$ much larger than the spin-relaxation length $\lambda_{\rm N}$. In this limit, the total correction to the current density from the combination of the SHE and ISHE is the sum of the corrections associated with the F-N interface at $z=0$ and the N-vacuum interface at $z=d_{\rm N}$, which is independent of the magnetization direction $\vm$. In this section we consider F$|$N bilayers and F$|$N$|$F trilayers with $d_{\rm N} \lesssim \lambda_{\rm N}$, for which the contributions from the two interfaces no longer add up.

For a normal layer of finite thickness $d_{\rm N}$, the ISHE correction $\delta \bar \vj_{\rm c}$ to the (effective) current densities follows by integration of Eqs.\ \eqref{eq:jcx1} and \eqref{eq:jcy1} over the entire cross section of the normal metal,
\begin{align}
  \delta \bar \vj_{\rm c}(\omega) =&\, 
    \frac{ \theta_{\rm SH} \sigma_{\rm N}}{2 e d_{\rm N}} \ve_z \times
    \left[ \vmu_{{\rm s}}(z \uparrow d_{\rm N},\omega) - \vmu_{{\rm s}}(z \downarrow 0,\omega) \right].
   \label{eq:deltaj_dN}
\end{align}
The $z$-dependence of the spin accumulation $\vmus$ is found from the solution of the spin-diffusion equation \eqref{eq:spindiffusion},
\begin{align}
  \vmus(z) =&\, \frac{1}{\sinh(d_{\rm N}/\lambda_{\rm N})}
  \left[ \vmus(z \downarrow 0)  \sinh \left( \frac{d_{\rm N} - z}{\lambda_{\rm N}} \right)
    \right. \nonumber \\ &\, \left. \mbox{}
    + \vmus(z \uparrow d_{\rm N})  \sinh \left( \frac{z}{\lambda_{\rm N}} \right) \right].
\end{align}
For the spin current densities $\vjs^z(z)$ at the normal-metal interfaces at $z=0$ and $z=d_{\rm N}$ one thus obtains, see Eq. (\ref{eq:jsz}), 
\begin{align}
  \vjs^z(0) =&\, \frac{1}{Z_{\rm N}} \left[\frac{\vmus(z \downarrow 0)}{\tanh(d_{\rm N}/\lambda_{\rm N})} - \frac{\vmus(z \uparrow d_{\rm N})}{\sinh(d_{\rm N}/\lambda_{\rm N})} \right] \nonumber \\ &\, \mbox{} - \theta_{\rm SH} \frac{\hbar \sigma_{\rm N} }{2 e} E \ve_y,
   \nonumber \\
  \vjs^z(d_{\rm N}) =&\, \frac{1}{Z_{\rm N}} \! \left[\frac{\vmus(z \downarrow 0)}{\sinh (d_{\rm N}/\lambda_{\rm N})} - \frac{\vmus(z \uparrow d_{\rm N})}{\tanh(d_{\rm N}/\lambda_{\rm N})} \right] \nonumber \\ &\, \mbox{}- \theta_{\rm SH} \frac{\hbar \sigma_{\rm N}}{2 e} E \ve_y,\label{eq:solution_N_spin_currents}
\end{align}
where the impedance $Z_{\rm N}$ is defined in Eq.\ (\ref{eq:ZN}).

In the limit $d_{\rm N} \gg \lambda_{\rm N}$, the terms proportional to $1/\sinh(d_{\rm N}/\lambda_{\rm N})$ can be neglected and the spin accumulations at the two interfaces at $z=0$ and $z=d_{\rm N}$ can be calculated separately. The resulting current correction is the sum of contributions from the two interfaces, as discussed above. To calculate the spin accumulations for finite $d_{\rm N}$, we now consider the cases of an F$|$N bilayer and an F$|$N$|$F trilayer separately.

\subsection{SMR for F$|$N bilayer}
\label{subsec:SMR_FN_bilayer}

The boundary condition for the insulating interface at $z = d_{\rm N}$ is that of zero spin current density,
\begin{equation}
  \vjs^z(d_{\rm N}) = 0.
\end{equation}
With the help of this boundary condition, the spin accumulation $\vmus(z \uparrow d_{\rm N})$ at the insulating surface may be eliminated from Eq.\ (\ref{eq:solution_N_spin_currents}). This allows us to express the spin current density $\vjs^z(0)$ at the ferromagnet interface at $z = 0$ in terms of the spin accumulation $\vmus(z \downarrow 0)$ at that interface,
\begin{align}
  \vjs^z(0) =&\, \frac{\vmus(z \downarrow 0)}{Z_{\rm N}}
  \tanh(d_{\rm N}/\lambda_{\rm N}) \nonumber \\ &\, \mbox{}
  - \theta_{\rm SH} \frac{\hbar \sigma_{\rm N}}{2 e} E \ve_y 
  \left[1 - \frac{1}{\cosh(d_{\rm N}/\lambda_{\rm N})} \right].
\end{align}
This is the same expression as Eq.\ (\ref{eq:ZN1}), but with the replacements
\begin{align}
  Z_{\rm N} \to&\, \tilde Z_{\rm N} \equiv
    Z_{\rm N} \coth(d_{\rm N}/\lambda_{\rm N}), \nonumber \\
  E \to &\,
  \tilde E = E \left[1 - \frac{1}{\cosh(d_{\rm N}/\lambda_{\rm N})} \right].
 \label{eq:Zsubst}
\end{align}
When these replacements are made, the calculation of \mbox{$\vmu_{\rm s}(z \downarrow 0)$} then follows that of Sec.\ \ref{sec:trans+long_SMR}. Compared to Sec.\ \ref{sec:trans+long_SMR}, the calculation of the charge currents involves the replacement of the spin accumulation $\vmu_{\rm s}(z \downarrow 0)$ at the F-N interface by the difference $\vmu_{\rm s}(z \downarrow 0) - \vmu_{\rm s}(z \uparrow d_{\rm N})$, see Eq.\ (\ref{eq:deltaj_dN}). Using that $j_{\rm s}^z(d_{\rm N}) = 0$, the spin accumulation $\vmu_{\rm s}(z \uparrow d_{\rm N})$ at the insulating surface can be easily obtained from Eq.\ (\ref{eq:solution_N_spin_currents}) and one finds that, up to a constant term that does not depend on the magnetization orientation, $\vmu_{\rm s}(z \downarrow 0) - \vmu_{\rm s}(z \uparrow d_{\rm N}) = (\tilde E/E) \vmu_{\rm s}(z \downarrow 0)$. As a consequence, the resulting charge currents $\delta \bar j_{\rm c}^x$ and $\delta \bar j_{\rm c}^y$ are still given by Eqs.\ (\ref{eq:jcx2}) and (\ref{eq:jcy2}), respectively, but with the substitution (\ref{eq:Zsubst}) for $Z_{\rm N}$, the substitution $E \to \tilde E^2/E$ for $E$, and with the corresponding substitutions $Z_{\parallel} \to \tilde Z_{\parallel}$ and $Z_{\perp} \to \tilde Z_{\perp}$ for the sum impedances $\tilde Z_{\parallel}$ and $\tilde Z_{\perp}$, see Eq.\ (\ref{eq:Ztotal}).

In the limit $d_{\rm N} \gg \lambda_{\rm N}$, the results of Sec.\ \ref{sec:trans+long_SMR} are recovered, because then $\tilde Z_{\rm N} \to Z_{\rm N} $ and $\tilde E \to E$. The leading correction term is an over-all factor $(1-4 e^{-d_{\rm N}/\lambda_{\rm N}})$ for the charge current $\delta \bar \vj_{\rm c}$, which comes from the substitution $E \to \tilde E^2/E$. In the opposite limit $d_{\rm N} \ll \lambda_{\rm N}$, we note that the prefactor $\tilde Z_{\rm N} \tilde E^2/d_{\rm N} E = (Z_{\rm N} E/d_{\rm N}) \tanh^2(d_{\rm N}/2 \lambda_{\rm N}) \tanh(d_{\rm N}/\lambda_{\rm N})$ in Eqs.\ \eqref{eq:jcx2} and \eqref{eq:jcy2} is proportional to $(d_{\rm N}/\lambda_{\rm N})^2$, so that the SMR is strongly suppressed. A further suppression of the SMR occurs once $d_{\rm N}$ is so small that both sum impedances $Z_{\parallel}$ and $Z_{\perp}$ of Eq.\ (\ref{eq:Ztotal}) are dominated by $\tilde Z_{\rm N}$. In this limit longitudinal and transverse contributions cancel and no observable SMR remains. This effect is not relevant for YIG$|$Pt, since it would require an unrealistically small layer thickness $d_{\rm N}$ for that material combination, but is relevant for Fe$|$Au, where this asymptotic small-$d_{\rm N}$ behavior sets in for $d_{\rm N} \lesssim 6 \cdot 10^{-8} \, \rm{m}$.

\subsection{SMR for F$|$N$|$F trilayer}
\label{subsec:SMR_FNF}

To describe the SMR in an F$|$N$|$F trilayer, we introduce two sets of spin impedances $Z_{{\rm F}\parallel}^{{\rm e}(j)}$, $Z_{{\rm F}\parallel}^{{\rm m}(j)}$, $Z_{{\rm F}\perp}^{(j)}$, $Z_{{\rm FN}\parallel}^{{\rm e}(j)}$, $Z_{{\rm FN}\parallel}^{{\rm m}(j)}$, and $Z_{{\rm FN}\perp}^{(j)}$ with $j=1,2$, where the superscripts $j=1$ and $j=2$ refer to the ferromagnet and the F-N interface at $z=0$ and $z=d_{\rm N}$, respectively, and the corresponding sum impedances $Z_{\parallel}^{(j)}$ and $Z_{\perp}^{(j)}$ are defined as in Eq.\ (\ref{eq:Ztotal}). The longitudinal and transverse components of the spin impedances are defined with respect to the unit vectors $\ve_{\parallel}^{(1)}$ and $\ve_{\parallel}^{(2)}$ pointing along the magnetization direction in the two magnets. Restoring the vector notation for the spin current density $\vj_{\rm s}^z(z)$ and the spin accumulation $\vmu_{\rm s}(z)$, the boundary conditions \eqref{eq:ZF1} and \eqref{eq:ZFN1} for spin currents and spin accumulations at the interfaces at $x=0$ and $x=d_{\rm N}$ can be summarized as
\begin{align}
  \vmus(z \downarrow 0) &=  - (\vZ^{(1)} - Z_{\rm N})
   \vjs^z(0), \nonumber\\
   \vmus(z \uparrow d_{\rm N}) &= + (\vZ^{(2)} - Z_{\rm N})
   \vjs^z(d_{\rm N}), \label{eq:mu_0_d_N}
\end{align}
where, for $j=1,2$,
\begin{align}
  \vZ^{(j)} =&\, Z_{\parallel}^{(j)} \ve_{\parallel}^{(j)} \ve_{\parallel}^{(j){\rm T}} 
  + Z_{\perp}^{(j)}(\omega) \ve_{\perp}^{(j)} \ve_{\perp}^{(j){\rm T}*}
  \nonumber \\ &\, \mbox{}
  + Z_{\perp}^{(j)*}(-\omega) \ve_{\perp}^{(j)*} \ve_{\perp}^{(j){\rm T}},
\end{align}
with ${\rm T}$ denoting the transpose vector and $*$ complex conjugation. The system of equations is closed by the boundary condition \eqref{eq:solution_N_spin_currents}. Solving for the spin accumulations $\vmus(z \downarrow 0)$ and $\vmus(z \uparrow d_{\rm N})$, one finds the charge current densities $\delta \bar j^{x,y}_{\rm c}$ from Eq.\ (\ref{eq:deltaj_dN}),
\begin{align}
  \label{eq:deltajcx_FNF}
  \delta \bar \vj_{\rm c}(\omega) =&\, 
  - \frac{\theta_{\rm SH}^2 \hbar \sigma_{\rm N}^2}{4 e^2 d_{\rm N}} E(\omega) Z_{\rm N}^2 \tanh^2 \frac{d_{\rm N}}{2 \lambda_{\rm N}}
  \\ \nonumber &\, \mbox{} \times
  \sum_{j=1}^{2} \left[ \ve_z \times 
  \vZ^{(j)}(\omega)^{-1} \vC(\omega)^{-1} \ve_y \right].
\end{align}
Here
\begin{align}
  \hat \vZ^{(j)} = Z_{\rm N} \coth \frac{d_{\rm N}}{2 \lambda_{\rm N}} \mI 
  + \vZ^{(j)},\ \ j=1,2,
  \label{eq:Zhat}
\end{align}
with ${\cal I}$ the identity matrix, and
\begin{align}
  \vC = \frac{1}{2} 
  \sum_{j} \check \vZ^{(j)} \hat \vZ^{(j)-1},
\end{align}
with
\begin{align}
  \check \vZ^{(j)} = Z_{\rm N} \tanh \frac{d_{\rm N}}{2 \lambda_{\rm N} }\mI 
  + \vZ^{(j)},\ \ j=1,2.
  \label{eq:Zcheck}
\end{align}
In Eq.\ (\ref{eq:deltajcx_FNF}) we subtracted a constant term that does not depend on the magnetization direction.
One verifies that $\vC(\omega) \to 1$ in the limit $d_{\rm N} \gg \lambda_{\rm N}$, so that the result for the charge current correction $\delta \bar \vj_{\rm c}$ is the sum of contributions from the two interfaces separately. The case of an F$|$N bilayer discussed in Subsec.\ \ref{subsec:SMR_FN_bilayer} can be obtained from Eq.\ (\ref{eq:deltajcx_FNF}) upon taking the limit $\vZ_{\rm FN}^{(2)} \to \infty$.

In the limit $d_{\rm N} \gg \lambda_{\rm N}$, one may approximate
\begin{equation}
  \vC^{-1} \approx \mI + 2 Z_{\rm N}
  e^{-d_{\rm N}/\lambda_{\rm N}} \sum_{j=1}^{2} \hat \vZ^{(j)-1}.
\end{equation}
In this limit, we find that Eq.\ (\ref{eq:deltajcx_FNF}) gives $\delta \bar \vj^{\rm c}$ as the sum of separate contributions from the two F layers in the form of Eqs. \eqref{eq:jcx2} and \eqref{eq:jcy2} and a weak interaction between the current corrections from the two interfaces,
\begin{widetext}
  \begin{align}
  \label{eq:jccorrection1}
  \delta \bar \vj_{\rm c}(\omega) =&\, [\delta \bar \vj_{\rm c}^{(1)}(\omega) + \delta \bar \vj_{\rm c}^{(2)}(\omega)](1 - 4 e^{-d_{\rm N}/\lambda_{\rm N}})
    \nonumber \\ &\, 
    - \frac{\theta_{\rm SH}^2 \hbar \sigma_{\rm N}^2}{2 e^2 d_{\rm N}} E(\omega) Z_{\rm N}^3 e^{d_{\rm N}/\lambda_{\rm N}}
    \ve_z \times [\hat \vZ^{(1)-1}(\omega) \hat \vZ^{(2)-1}(\omega) + \hat \vZ^{(2)-1}(\omega) \hat \vZ^{(1)-1}(\omega)] \ve_y,
\end{align}
\end{widetext}
where $\delta \bar \vj_{\rm c}^{(j)}$ is the correction to the charge current for a single F-N interface, $j=1,2$, and the factor \mbox{$(1-4e^{-d_{\rm N}/\lambda_{\rm N}})$} is the leading correction factor for a finite-width interface, see the discussion below Eq.\ (\ref{eq:Zsubst}).

In the opposite limit $d_{\rm N} \ll \lambda_{\rm N}$, one has
\begin{align}
  \vC^{-1} \approx \frac{4 \lambda_{\rm N} Z_{\rm N}}{d_{\rm N}}
  \vZ_{\Sigma}^{-1} - 2 Z_{\rm N} \vZ_{\Sigma}^{-1} (2 Z_{\rm N} - \vZ_{\Sigma}') 
  \vZ_{\Sigma}^{-1},
\end{align}
where we abbreviated 
\begin{align}
  \vZ_{\Sigma} = \sum_{j=1}^{2} (\vZ_{\rm FN}^{(j)} + \vZ_{\rm F}^{(j)}),\ \
  \vZ_{\Sigma}' = \frac{1}{Z_{\rm N}} \sum_{j=1}^{2} (\vZ_{\rm FN}^{(j)} + \vZ_{\rm F}^{(j)})^2.
\end{align}
In the same manner, the sum of inverses
\begin{equation}
  \sum_{j}\vZ^{(j)-1} \approx \frac{d_{\rm N}}{\lambda_{\rm N} Z_{\rm N}} \mI
  - \frac{d_{\rm N}^2}{4 \lambda_{\rm N}^2 Z_{\rm N}^2} \vZ_{\Sigma}.
\end{equation}
Combining these results, we find that for small $d_{\rm N}/\lambda_{\rm N}$ one has
\begin{align}
  \label{eq:jccorrection2}
  \delta \bar \vj_{\rm c}(\omega) =&\, 
  \frac{4 e^2 \theta_{\rm SH}^2 d_{\rm N}}{\hbar} E(\omega) \ve_{z} \times \left[ \vZ_{\Sigma}^{-1} 
    \vphantom{\frac{M}{M}} \right.
  \\ \nonumber &\, 
  \left. \mbox{} - \frac{d_{\rm N}}{2 \lambda_{\rm N}} 
  \vZ_{\Sigma}^{-1} (2 Z_{\rm N} - \vZ_{\Sigma}') \vZ_{\Sigma}^{-1} \right]  \ve_y, 
\end{align}  
where we omitted a constant term that does not depend on the magnetization direction.

\begin{figure}
\centering
\includegraphics[width=0.5\textwidth]{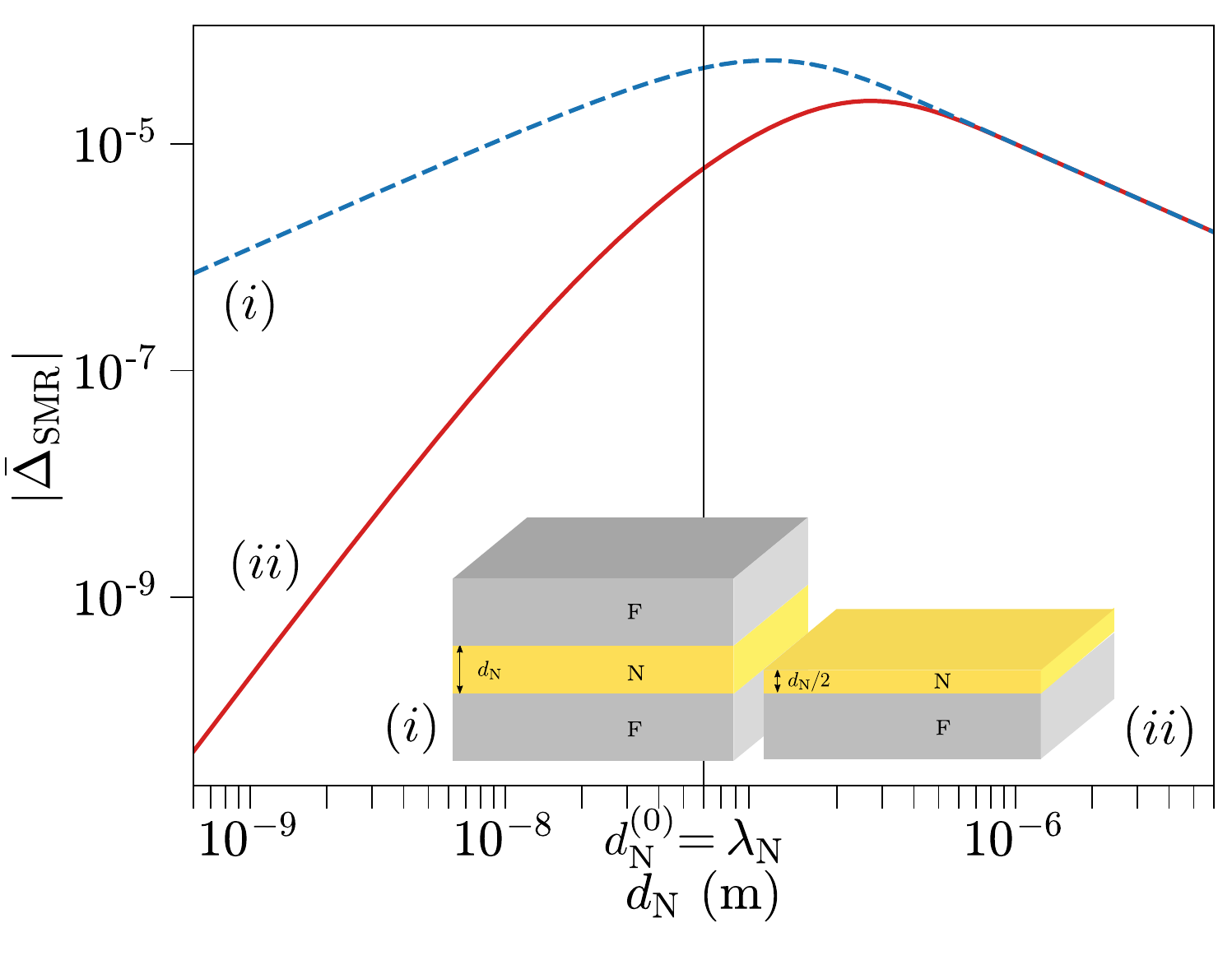}
\caption{Dimensionless characteristic $\bar \Delta_{\rm SMR}$, see Eq.\ (\ref{eq:DeltaDef}), of a Fe$|$Au$|$Fe trilayer (dashed blue line) and a Fe$|$Au bilayer (solid red line) as function of the thickness of the Au layer. The Fe$|$Au$|$Fe trilayer has identical F layers with parallel magnetization directions and a thickness $d_{\rm N}$, while the thickness of the Au layer in the Fe$|$Au bilayer is $d_{\rm N}/2$. (This ensures that the magnetic interface/volume ratio in both geometries is same.) Parameter values are taken from Tabs.\ \ref{tab:estimates_parameters} and \ref{tab:estimates_impedances}. The two curves are for $\omega/2\pi = 0$\,Hz. For comparison, $d_{\rm N}^{(0)} = 6 \cdot 10^{-8}$\,m is the thickness of the N layer of Ref.\ \onlinecite{Urban_2001}, which is also the reference thickness used for the numerical estimates of Sec.\ \ref{sec:experiments}.
\label{fig:YIG|Pt_SMR_ratios_d_N}}
\end{figure}

In the limit of a small layer width $d_{\rm N}$, the charge current correction $\delta \bar \vj_{\rm c}$ of Eq.\ (\ref{eq:jccorrection2}) for an F$|$N$|$F trilayer is a factor $(d_{\rm N}/\lambda_{\rm N})^2$ smaller than the SMR for two independent F-N interfaces. To understand why, note that for small $d_{\rm N}/\lambda_{\rm N}$ the coupled equations (\ref{eq:solution_N_spin_currents}) imply that both the spin currents $\vj_{\rm s}^z$ and the spin accumulations at the interfaces are proportional to $d_{\rm N}/\lambda_{\rm N}$. This reflects the fact that the interface spin accumulation generated by the spin-Hall effect is proportional to the layer thickness $d_{\rm N}$ for small $d_{\rm N}$. However, to this order in the layer thickness $d_{\rm N}$, the difference $\vmu_{\rm s}(z \downarrow 0) - \vmu_{\rm s}(z \uparrow d_{\rm N})$ of the spin accumulations, which is what determines the charge current correction, see Eq.\ (\ref{eq:deltaj_dN}), depends on the driving field $E$ and the spin-Hall angle $\theta_{\rm SH}$, but not on the orientations of the magnetizations of the two F layers. This can be seen, {\it e.g.}, by taking the sum of the two equations in Eq.\ (\ref{eq:solution_N_spin_currents}), which gives $\vmu_{\rm s}(z \downarrow 0) - \vmu_{\rm s}(z \uparrow d_{\rm N}) = 2 e d_{\rm N} \theta_{\rm SH} E \ve_y$ to leading (first) order in $d_{\rm N}$. The dependence on the magnetization direction, which is what constitutes the SMR, occurs to subleading (second) order in $d_{\rm N}/\lambda_{\rm N}$, which explains the smallness of the charge current by a factor $(d_{\rm N}/\lambda_{\rm N})^2$ in the small-$d_{\rm N}$ limit.

In spite of its smallness, the charge current correction $\delta \vj_{\rm c}$ for an F$|$N$|$F trilayer is a factor $\sim \lambda_{\rm N}/d_{\rm N}$ larger than the charge current correction for an F$|$N bilayer, which was discussed in Subsec.\ \ref{subsec:SMR_FN_bilayer}. The reason is that for the F$|$N bilayer the spin current $\vj_{\rm s}^z(d_{\rm N})$ is strictly zero, whereas for the F$|$N$|$F trilayer one ferromagnet can serve as a ``sink'' for the spin current generated by the other ferromagnet and vice versa, so that both spin currents $\vj_{\rm s}^z(0)$ and $\vj_{\rm s}^{z}(d_{\rm N})$ can be nonzero. This observation was made by Chen {\it et al.} for the DC limit.\cite{Chen_2013} Our calculation shows that the same mechanism also applies at finite driving frequency.

Fig. \ref{fig:YIG|Pt_SMR_ratios_d_N} shows the characteristic $\bar \Delta_{\rm SMR}$ of the spin-Hall magnetoresistance in an Fe$|$Au bilayer and an Fe$|$Au$|$Fe trilayer with Au layers of thickness $d_{\rm N}/2$ and $d_{\rm N}$, respectively, as a function of $d_{\rm N}$ for $\omega = 0$. The magnetization directions $\ve_{\parallel}^{(1)}$ and $\ve_{\parallel}^{(2)}$ in the F$|$N$|$F trilayer are parallel. In the large-$d_{\rm N}$ limit, $\bar \Delta_{\rm SMR}$ is the same for both configurations, because in this limit the two F layers of the F$|$N$|$F trilayer are independent and their identical contributions simply add up. For large $d_{\rm N}$, the characteristic $\Delta_{\rm SMR}^{\infty}$ is inversely proportional to $d_{\rm N}$, since the magnetization-direction dependent contribution to the conductivity exists on top of a large background current independent of the magnetization direction. For small $d_{\rm N}$, $\bar \Delta_{\rm SMR}$ differs for a bilayer and a trilayer because of the interplay between the two ferromagnetic layers; $\bar \Delta_{\rm SMR} \propto d_{\rm N}^2$ for the F$|$N bilayer, which is the asymptotic dependence discussed in Subsec.\ \ref{subsec:SMR_FN_bilayer}, whereas $\bar \Delta_{\rm SMR} \propto d_{\rm N}$ for the F$|$N$|$F trilayer. 

\subsection{F$|$N$|$F trilayer with collinear magnetization directions}
\label{subsec:parallel_spin_valve_config}

If the magnetization directions $\ve_{\parallel}^{(1)}$ and $\ve_{\parallel}^{(2)}$ are collinear, {\it i.e.}, $\ve_{\parallel}^{(1)} = \pm \ve_{\parallel}^{(2)}$, there exists a basis in which the matrices $\vZ^{(1)}$ and $\vZ^{(2)}$ can be diagonalized simultaneously. In this limit, it is possible to directly express the charge current components $\delta \bar j_{\rm c}^x$ and $\delta \bar j_{\rm c}^y$ in terms of the scalar impedances $Z_{\rm N}^{(j)}$, $Z_{\rm FN}^{(j)}$, and $Z_{\rm F}^{(j)}$, $j=1,2$. For the case of parallel magnetization directions, $\ve_{\parallel}^{(1)} = \ve_{\parallel}^{(2)}$, one finds (up to constant terms that do not depend on the magnetization direction)
\begin{widetext}
\begin{align}
\label{eq:deltajcx_FNF_parallel}
  \delta \bar j_{\rm c}^{x}(\omega) =&\, - \frac{\theta^2_{\rm SH} \hbar \sigma_{\rm N}^2}{2 e^2 d_{\rm N}} E(\omega) Z_{\rm N}^2 \tanh^2 \frac{d_{\rm N}}{2 \lambda_{\rm N}}
  \left[ \frac{1}{2} (1-m_y^2) \frac{ \hat{Z}_{\perp}^{(1)*}(- \omega) + \hat{Z}^{(2)*}_{\perp}(-\omega)}{ \check{Z}^{(1)*}_{\perp}(-\omega) \hat{Z}_{\perp}^{(2)*}(- \omega) + \check{Z}^{(2)*}_{\perp}(- \omega) \hat{Z}^{(1)*}_{\perp}(-\omega) }
  \right. \\ &\,  \left. \nonumber \mbox{}
  + \frac{1}{2} (1-m_y^2) \frac{ \hat{Z}^{(1)}_{\perp}( \omega) + \hat{Z}^{(2)}_{\perp}(\omega)}{ \check{Z}^{(1)}_{\perp}(\omega) \hat{Z}^{(2)}_{\perp}( \omega) + \check{Z}^{(2)}_{\perp}( \omega) \hat{Z}^{(1)}_{\perp}(\omega) }
  + m_y^2 \frac{\hat{Z}_{\parallel}^{(1)} + \hat{Z}_{\parallel}^{(2)}}{\check{Z}_{\parallel}^{(1)} \hat{Z}_{\parallel}^{(2)}  + \check{Z}_{\parallel}^{(2)} \hat{Z}_{\parallel}^{(1)} } \right], \nonumber\\
\label{eq:deltajcy_FNF_parallel}
\delta \bar j_{\rm c}^{y}(\omega) =&\, -\frac{\theta^2_{\rm SH} \hbar \sigma_{\rm N}^2}{2 e^2 d_{\rm N}} E(\omega) Z_{\rm N}^2 \tanh^2 \frac{d_{\rm N}}{2 \lambda_{\rm N}}
\left[
 \frac{1}{2} (m_x m_y - i m_z)
  \frac{ \hat{Z}_{\perp}^{(1)*}(- \omega) + \hat{Z}^{(2)*}_{\perp}(-\omega)}{ \check{Z}^{(1)*}_{\perp}(-\omega) \hat{Z}_{\perp}^{(2)*}(- \omega) + \check{Z}^{(2)*}_{\perp}(- \omega) \hat{Z}^{(1)*}_{\perp}(-\omega) }
 \right. \\ &\, \left. \mbox{}
  + \frac{1}{2} (m_x m_y + i m_z)
   \frac{ \hat{Z}^{(1)}_{\perp}( \omega) + \hat{Z}^{(2)}_{\perp}(\omega)}{ \check{Z}^{(1)}_{\perp}(\omega) \hat{Z}^{(2)}_{\perp}( \omega) + \check{Z}^{(2)}_{\perp}( \omega) \hat{Z}^{(1)}_{\perp}(\omega) }
   - m_x m_y \frac{\hat{Z}_{\parallel}^{(1)} + \hat{Z}_{\parallel}^{(2)}}{\check{Z}_{\parallel}^{(1)} \hat{Z}_{\parallel}^{(2)}  + \check{Z}_{\parallel}^{(2)} \hat{Z}_{\parallel}^{(1)} }
   \right]. \nonumber
\end{align}
Here $\hat{Z}^{(j)}_{\perp,\parallel}( \omega)$ and $\check{Z}^{(j)}_{\perp,\parallel}( \omega)$ are the sum impedances of layer $j$, $j=1,2$, see Eq.\ (\ref{eq:Ztotal}), with $Z_{\rm N}$ replaced by $Z_{\rm N} \coth(d_{\rm N}/2 \lambda_{\rm N})$ and $Z_{\rm N} \tanh(d_{\rm N}/2 \lambda_{\rm N})$, respectively (compare with Eqs.\ (\ref{eq:Zhat}) and (\ref{eq:Zcheck})).
For the antiparallel configuration $\ve_{\parallel}^{(1)} = -\ve_{\parallel}^{(2)}$ the charge current correction $\delta \bar j_{\rm c}^{x,y}$ can be obtained from Eqs.\ (\ref{eq:deltajcx_FNF_parallel}) and (\ref{eq:deltajcy_FNF_parallel}) by exchanging the impedances $\check{Z}^{(2)}_{\perp}(\omega)$ and $\hat{Z}^{(2)}_{\perp}( \omega)$ by $\check{Z}^{(2)*}_{\perp}(- \omega)$ and $\hat{Z}^{(2)*}_{\perp}(- \omega)$, respectively, and vice versa. This substitution rule follows from the observation that the roles of the vectors $\ve_{\perp}^{(2)}$ and $\ve_{\perp}^{(2)*}$ are interchanged upon inverting the magnetization direction $\ve^{(2)}_{\parallel}$.

In the DC limit and neglecting the imaginary part of the spin-mixing conductances at the F-N interfaces, the current correction $\delta \bar \vj_{\rm c}$ in an F$|$N$|$F trilayer is the same for the parallel and antiparallel magnetization configuration.\cite{Chen_2013} At finite frequencies, the resonant features of $\delta \bar \vj_{\rm c}$ at the magnon frequencies are different for the parallel and antiparallel magnetization configuration: Upon inverting the magnetization direction of one of the magnets, the polarization vector of a propagating magnon mode in that magnet changes from $\ve_{\perp}$ to $\ve_{\perp}^*$ or vice versa, which affects the interaction term involving both magnets. (Mathematically, the difference between parallel and antiparallel magnetization configurations at large frequencies follows, because $Z_{{\rm F}\perp}(\omega)$ shows resonant features for positive frequencies, but not for negative frequencies.)

\subsection{SMR in the perpendicular F$|$N$|$F configuration}
\label{subsec:perp_spin_valve_config}

When the spin valve is in the perpendicular configuration, $\ve_{\parallel}^{(1)} \cdot \ve_{\parallel}^{(2)} = 0$, a simultaneous diagonalization of the matrices $\vZ^{(1)}$ and $\vZ^{(2)}$ is possible only in the DC limit $\omega \to 0$ and neglecting the imaginary part of the spin-mixing conductance at the F-N interfaces. In this limit, all impedances are real and one finds
\begin{align}
  \label{eq:jcperpx}
  \delta \bar j_{\rm c}^{x}(0) =&\, - \frac{\theta^2_{\rm SH} \hbar \sigma_{\rm N}^2}{2 e^2 d_{\rm N}} E(\omega) Z_{\rm N}^2 \tanh^2 \frac{d_{\rm N}}{2 \lambda_{\rm N}}
  \left[ m_y^{(1)2} \frac{\hat{Z}_{\parallel}^{(1)} + \hat{Z}_{\perp}^{(2)}}{\check{Z}_{\parallel}^{(1)} \hat{Z}_{\perp}^{(2)}  + \check{Z}_{\perp}^{(2)} \hat{Z}_{\parallel}^{(1)} } +  m_y^{(2)2} \frac{\hat{Z}_{\perp}^{(1)} + \hat{Z}_{\parallel}^{(2)}}{\check{Z}_{\perp}^{(1)} \hat{Z}_{\parallel}^{(2)}  + \check{Z}_{\parallel}^{(2)} \hat{Z}_{\perp}^{(1)}}
    \nonumber \right. \\ &\, \ \ \ \ \left. \mbox{}
  + (1- m_y^{(1)2} - m_y^{(2)2}) \frac{\hat{Z}_{\perp}^{(1)} + \hat{Z}_{\perp}^{(2)}}{\check{Z}_{\perp}^{(1)} \hat{Z}_{\perp}^{(2)}  + \check{Z}_{\perp}^{(2)} \hat{Z}_{\perp}^{(1)}}
    \right], \\
  \label{eq:jcperpy}
  \delta \bar j_{\rm c}^{y}(0) =&\, \frac{\theta^2_{\rm SH} \hbar \sigma_{\rm N}^2}{2 e^2 d_{\rm N}} E(\omega) Z_{\rm N}^2 \tanh^2 \frac{d_{\rm N}}{2 \lambda_{\rm N}}
  \left[ m_x^{(1)} m_y^{(1)} \frac{\hat{Z}_{\parallel}^{(1)} + \hat{Z}_{\perp}^{(2)}}{\check{Z}_{\parallel}^{(1)} \hat{Z}_{\perp}^{(2)}  + \check{Z}_{\perp}^{(2)} \hat{Z}_{\parallel}^{(1)} }
    +  m_x^{(2)} m_y^{(2)} \frac{\hat{Z}_{\perp}^{(1)} + \hat{Z}_{\parallel}^{(2)}}{\check{Z}_{\perp}^{(1)} \hat{Z}_{\parallel}^{(2)}  + \check{Z}_{\parallel}^{(2)} \hat{Z}_{\perp}^{(1)}}
    \nonumber \right. \\ &\, \ \ \ \ \left. \mbox{}
  - (m_x^{(1)} m_y^{(1)} + m_x^{(2)} m_y^{(2)}) \frac{\hat{Z}_{\perp}^{(1)} + \hat{Z}_{\perp}^{(2)}}{\check{Z}_{\perp}^{(1)} \hat{Z}_{\perp}^{(2)}  + \check{Z}_{\perp}^{(2)} \hat{Z}_{\perp}^{(1)}}
    \right].
\end{align}
\end{widetext}
This results generalizes an expression obtained in Ref.\ \onlinecite{Chen_2013}, which considers the case of two insulating ferromagnets and a perpendicular magnetization configuration with both magnetizations in the $xy$ plane. In this case the third terms between the square brackets in Eqs.\ (\ref{eq:jcperpx}) and (\ref{eq:jcperpy}) are independent of the magnetization direction. Moreover, since $Z^{(j)}_{\parallel}(\omega) \gg Z^{(j)}_{\perp}(\omega)$ for insulating ferromagnets at frequency $\omega \to 0$, each of the two remaining terms in the expressions for $\delta \bar j_{\rm c}^x$ and $\delta \bar j_{\rm c}^y$ depends on properties of a single magnet only. Indeed, if the magnetization directions $\ve_{\parallel}^{(1)}$ and $\ve_{\parallel}^{(2)}$ are both in the $xy$ plane and if the spin-mixing conductance of the F-N interface is real, there is nothing in the system that rotates spin currents $\vj^z_{\rm s}$ out of the $xy$ plane in the DC limit $\omega=0$. Since each ferromagnet drives only spin currents that are transverse to its magnetization direction, spin currents driven by one magnet are longitudinal for the other magnet and vice versa. Longitudinal spin currents are fully reflected by insulating ferromagnets in the DC limit, which explains why there is no ``interaction'' contribution to $\delta \vj_{\rm c}$ in this case. Our full result of Eqs.\ (\ref{eq:jcperpx}) and (\ref{eq:jcperpy}) shows that this is a special property of the case that both magnetizations are in the $xy$ plane and that the magnets are insulating. If either of these conditions is lifted, $\delta \bar \vj_{\rm c}$ contains interaction terms which depend on properties of both magnets.

At finite frequencies or if the imaginary part of the interface impedances is taken into account, it is no longer possible to simultaneously diagonalize $\vZ^{(1)}$ and $\vZ^{(2)}$. In this case, no simple closed-form expressions for the charge current correction $\delta \vj_{\rm c}$ could be obtained and one has to resort to the matrix expression (\ref{eq:deltajcx_FNF}) or its asymptotic limits (\ref{eq:jccorrection1}) and (\ref{eq:jccorrection2}) for large and small $d_{\rm N}/\lambda_{\rm N}$.

\end{appendix}

\bibliography{./ac-smr_theory_paper_final.bib}

\end{document}